\pdfoutput=1
\documentclass[journal,twoside]{IEEEtran}
\usepackage[utf8]{inputenc}
\usepackage{cite}
\usepackage{amsmath,amssymb,amsfonts}
\usepackage{algorithmic}
\usepackage{graphicx}
\usepackage{textcomp}
\usepackage[dvipsnames]{xcolor}
\usepackage{subfig}
\usepackage{float}
\usepackage{multirow}
\usepackage[explicit]{titlesec}
\usepackage{setspace}

\titlespacing{\section}{0pt}{1.2ex plus .0ex minus .0ex}{.3ex plus .0ex}
\titlespacing{\subsection}{0pt}{1.2ex plus .0ex minus .0ex}{.3ex plus .0ex}

\def\BibTeX{{\rm B\kern-.05em{\sc i\kern-.025em b}\kern-.08em
    T\kern-.1667em\lower.7ex\hbox{E}\kern-.125emX}}
\begin{document}

\title{
	A Dynamic Equivalent Method for PMSG-WTG \\
	 Based Wind Farms Considering Wind Speeds\\ and Fault Severities}
\author{Dongsheng Li,~\IEEEmembership{Student Member,~IEEE}, Chen Shen,~\IEEEmembership{Senior Member,~IEEE}, Ye Liu,~\IEEEmembership{Student Member,~IEEE}, 
	
	Ying Chen,~\IEEEmembership{Senior Member,~IEEE} and Shaowei Huang,~\IEEEmembership{Member,~IEEE}
	\thanks{This work is supported by the National Natural Science Foundation of China under Grant U2166601. (Corresponding author: Chen Shen.)

Dongsheng Li, Chen Shen, Ye Liu, Ying Chen and Shaowei Huang are with the State Key Laboratory of Power Systems, Department of Electrical Engineering, Tsinghua University, Beijing 100084, China (e-mail: lids19@mails.tsinghua.edu.cn; shenchen@mail.tsinghua.edu.cn; liuye18@mails.tsinghua.edu.cn; chen\_ ying@tsinghua.edu.cn;
huangsw@tsinghua.edu.cn).
}}

\markboth{Journal of \LaTeX\ Class Files,~Vol.~XX, No.~XX, XX~XXXX}%
{Journal of \LaTeX\ Class Files,~Vol.~XX, No.~XX, XX~XXXX}

\maketitle

\begin{abstract}
The dynamic security assessment of power systems needs to scan contingencies in a preselected set through time-domain simulations. With more and more inverter-based-generation, such as wind and solar power generation, integrated into power systems, electro-magnetic transient simulation is adopted. However, the complexity of simulation will increase greatly if inverter-based-generation units are modeled in detail. In order to reduce the complexity of simulation of power systems including large-scale wind farms, it is critical to develop dynamic equivalent methods for wind farms which are applicable to the expected contingency analysis. The dynamic response characteristics of permanent magnet synchronous generator-wind turbine generators (PMSG-WTGs) are not only influenced by their control strategies, but also by the operating wind speeds and the fault severities. Thus, this paper proposes a dynamic equivalent method for PMSG-WTG based wind farms considering the wind speeds and the fault severities. Firstly, this paper analyzes all possible response characteristics of a PMSG-WTG and proposes a clustering method based on the operating wind speed and the terminal voltage of each PMSG-WTG at the end of the fault. Then, a single-machine equivalent method is introduced for each group of PMSG-WTGs. For the group of PMSG-WTGs with active power ramp recovery process, an equivalent model with segmented ramp rate limitation for active current is designed. In order to obtain the clustering indicators, a simulation-based iterative method is put forward to calculate the voltage at point of common connection (PCC) of the wind farm, and a PMSG-WTG terminal voltage calculation method is further presented. Eventually, the efficiency and accuracy of the proposed method are verified by the simulation results.
\end{abstract}

\setstretch{0.93} 
\section{Introduction}
With its safety, cleanliness and high efficiency, wind power has been widely integrated in power systems \cite{mahela2016comprehensive}. However, large-scale wind power integration has great effects on the stable operation of power systems due to the randomness and fluctuation of wind power \cite{wang2019approaches}. In order to maintain the reliability of power supply, it is necessary to perform the dynamic security assessment (DSA) of power systems with large-scale wind power integration, which is always realized by the time domain simulation methods. However, if modeling every wind power generator in detail during simulation, the computational complexity will be increased greatly, and even the problem of “dimension disaster” will occur \cite{zou_survey_2014}. In addition, the DSA of power systems needs to be applied to the expected contingencies, so that preventive measures can be taken to ensure the system safety even if the expected contingencies occur \cite{__2004}. Therefore, it is of vital significance to establish a dynamic equivalent model of wind farm which is suitable for analyzing the expected contingencies.

The existing methods for dynamic equivalent modeling are divided into two categories. The first category is the single-machine equivalent method, which can be further classified into one wind turbine with one generator model and multiple wind turbines with one generator model. In the former model, the equivalent wind speed is calculated by different methods, such as the wind energy utilization coefficient weighting method \cite{__2013} and the equivalent methods based on constant active power \cite{trudnowski_fixed-speed_2004,brochu_validation_2011,qi_dynamic_2016}. Other parameters of the equivalent model are obtained by the capacity weighting method. Moreover, there is literature using optimization algorithms to calculate the control parameters of the equivalent model, such as genetic learning particle swarm optimization algorithm \cite{zhang2020robustness} and recursive least-squares method \cite{kim_dynamic_2016}, which make the dynamic response of the equivalent model closer to the reality. In the latter model, the sum of mechanical torques of each wind turbine is set as the input mechanical torque of equivalent model \cite{mercado-vargas_aggregated_2015,fernandez_aggregated_2008}. The single-machine equivalent method is effective when the wind speeds in the equivalently modeled wind farm have no obvious difference. However, with the continuous increase of the scale of wind farms, the operating wind speeds of wind turbines in the same wind farm vary a lot. Thus, the single-machine equivalent method cannot represent the dynamic characteristics of a whole wind farm accurately.

The second category is the multi-machine equivalent method, which is able to represent the different behaviors among wind power generators. In the method, one or a group of features that can characterize the operating state of wind power generators are usually selected as the clustering indicators. However, these methods sometimes require the post-fault information, such as the reactive power of wind generator \cite{_bp_2019}, the rotor currents \cite{__2015-2,han_novel_2022} or the stator short-circuit currents \cite{zou_fuzzy_2015}, which are suitable for post-fault analysis, but cannot analyze the expected contingencies. There are also multi-machine equivalent methods that only use pre-fault information, such as the wind speed \cite{han2021real,__2015,li_practical_2018}, the pitch angle \cite{li2020research,__2010} and the active power sequence before the fault \cite{__2015-1}, which are able to analyze the expected contingencies. Nevertheless, wind power generators are clustered without considering the influence of the fault severity in \cite{han2021real,__2015,li_practical_2018,li2020research,__2010,__2015-1}. Considering the reactive power priority control and the active power ramp recovery process, the pre-fault states of wind power generators and the fault severity together determine the dynamic responses of wind power generators. Thus, wind power generators cannot be clustered correctly under different faults using the methods proposed in \cite{han2021real,__2015,li_practical_2018,li2020research,__2010,__2015-1}. In \cite{wang2020dynamic}, wind power generators are clustered by the similarity among the measurable output characteristics under a specific fault. However, the results of clustering may not be applicable to other fault conditions. In \cite{_crowbardfig_2015,ding_equivalent_2016}, wind power generators are divided into different clusters according to the operation characteristics of crowbars, which are influenced by the terminal voltages of wind power generators. The terminal voltages are assumed as constants during a fault in \cite{_crowbardfig_2015,ding_equivalent_2016}. However, the assumption is unreasonable while wind power generators will output reactive power to support the terminal voltages during a fault. In addition, the voltage of dc-link capacitor cannot exactly reflect the dynamic response characteristics of wind power generators when considering the reactive power priority control and active power ramp recovery process. 

The active power of wind power generators operating at different wind speeds will restore to their pre-fault value with different recovery time after the fault clearance. If the wind farm is clustered by the traditional single-machine equivalent method, the active power of the equivalent model will restore to its pre-fault active power at a certain ramp rate after the fault clearance, which is inaccurate during the post-fault recovery process. In \cite{li_practical_2018}, a four-machine equivalent method is proposed to fit the different recovery characteristics of wind power generators with four different recovery rates. When a large number of wind power generators are under consideration, the active power recovery process is divided into many segments with different recovery rates which is far more than four segments. Thus, the method is still inaccurate when considering large scale wind farms. In \cite{chao2021analytical,li2017improved}, the reference active power of each doubly fed induction generator (DFIG) after the fault clearance is calculated analytically by the wind speeds and the terminal voltages. And the reference active power of the equivalent model is obtained by the sum of the reference active power of all DFIGs. Nevertheless, as for the permanent magnet synchronous generator-wind turbine generator (PMSG-WTG), the reference value of active current is influenced by the characteristic of the dc-link voltage, which cannot be calculated analytically. Moreover, the existence of the ramp recovery characteristics also depends on both the wind speed and the severity of fault when considering the reactive power priority control and the active power ramp recovery process. Thus, the methods proposed in \cite{chao2021analytical,li2017improved} are not applicable to the PMSG-WTG.

In summary, it is difficult to find a dynamic equivalent method that considers both the pre-fault operating state and the severity of fault when takes account of the reactive power priority control and active power ramp recovery process. Therefore, this paper try to develop an equivalent method for PMSG-WTG based wind farms, which takes the influence of wind speed, the fault severity, the reactive power priority
control and the active power ramp recovery process into account.  
The main contributions are listed as follows:
\begin{itemize}
\item[1)] A clustering method is proposed considering pre-fault wind speeds and PMSG-WTG terminal voltages at the end of a fault. All possible dynamic response characteristics of PMSG-WTGs and the corresponding boundary conditions are derived considering the reactive power priority control and the active power ramp recovery process. Moreover, a convenient clustering method based on wind speed and terminal voltage is put forward.
\item[2)] A single-machine equivalent model is proposed for each subgroup of PMSG-WTGs. However, the model is still cannot reflect the difference  among PMSG-WTGs with active power ramp recovery process. Thus, an improved single-machine equivalent model with segmented ramp rate limitation for active current is presented for the subgroup.
\item[3)] In order to obtain the clustering indicators of each PMSG-WTG, a simulation-based iterative method is introduced to calculate the voltage at point of common connection (PCC) of the wind farm to be equivalently modeled at the end of faults. And a terminal voltage calculation method is further designed. Based on the voltage calculation method, the proposed equivalent method can be applied to any expected contingencies.
\end{itemize}

The rest of the paper is arranged as follows: the control strategy of PMSG-WTG is introduced in Section II. A clustering method considering wind speeds and terminal voltages is put forward in Section III. Different single-machine equivalent methods are designed for each subgroup of PMSG-WTGs in Section IV. A terminal voltage calculation method is presented in Section V. The proposed equivalent method is verified with different faults in IEEE 39-bus system in Section VI. Conclusions are drawn in Section VII.
 
\section{Control Strategy of PMSG-WTG}

In the PMSG-WTG studied in this paper, the machine-side converter (MSC) consists of an uncontrolled diode rectifier bridge and a Boost chopper circuit. The grid-side converter (GSC) consists of a controlled inverter bridge composed of insulated gate bipolar transistors (IGBT). The control strategies of converters are described below. Models of other parts of the PMSG-WTG such as wind turbine, drive train system and synchronous machine are consistent with the conventional models, which can be found in the literature \cite{__2019}. 
\subsection{ Control Strategy of MSC}
The active power control strategy of MSC is to achieve the maximum power point trace (MPPT). The optimal reference rotor speed can be calculated by:
\begin{equation}\omega_{opt}=\lambda_{opt} V_{w}/\gamma \label{1}
\end{equation}
where $\omega_{opt}$ is the optimal mechanical angular speed of wind turbine; $V_{w}$ is the wind speed; $\gamma$ is the wind turbine radius;  $\lambda_{opt}$ is the optimal tip speed ratio, which is a constant in the MPPT control. 
The boost IGBT switching signal is generated using the duty ratio, which can be obtained by the PI controller with eliminating the difference between the real dc current and the reference dc current. 
\subsection{ Control Strategy of GSC}
The GSC adopts grid voltage-oriented vector control to stabilize the dc-link voltage and realize the unity power factor control. The $dq$-axis voltage of grid can be derived as:
\begin{equation}
\begin{cases}
u_{d}=e\\
u_{q}=0\label{eq2}
\end{cases}
\end{equation}
where $e$ is the magnitude of grid voltage space vector; $u_{d}$ and $u_{q}$ are the $d$-axis and $q$-axis voltage of grid,respectively.

The active and reactive output power of GSC are:
\begin{equation}
\begin{cases}
P=\frac{3}{2}ei_{d}\\
Q=\frac{3}{2}ei_{q}\label{eq3}
\end{cases}
\end{equation}
where $i_{d}$ and $i_{q}$ are the $d$-axis and $q$-axis currents of grid side, respectively. It can be found that the  active and reactive output power of GSC can be controlled independently by $i_{d}$ and $i_{q}$. Thus, $i_{d}$ and $i_{q}$ are also called active current and reactive current, respectively.
\subsubsection{Control Strategy of GSC During Normal Operation }
During normal operation, the dc-link voltage is compared with the reference voltage and the error is PI controlled to obtain the reference value of $d$-axis current. The reference value of reactive power is usually set to 0 to achieve the unity power factor control. Thus, the reference value of $q$-axis current is also 0.
\subsubsection{Control Strategies of GSC During and After a Fault}
The GSC is unable to output the active power normally when the grid-side voltage suddenly drops, which will lead to the increase of the dc-link voltage. The chopper circuit connected in parallel with the dc capacitor will operate and consume the excess active power when the dc-link voltage reaches the set threshold. Since the dc-link voltage is always higher than the reference value during a fault, the $d$-axis current will continuously increase until the dc-link voltage returns to its reference value or the $d$-axis current is limited by the converter capacity. At the same time, PMSG-WTGs are required to generate reactive power to support the grid voltage during a fault. According to the grid codes \cite{GB36995}, the reference value of output $q$-axis current during low-voltage ride-through (LVRT) in this paper is:
\begin{equation}I_{qref}=1.5\times (0.9-U_{T})I_{N},(0.2\le U_{T} \le 0.9)\label{eq5}
\end{equation}
where $I_{qref}$ is the reference value of $q$-axis current; $U_{T}$ is the terminal voltage of PMSG-WTG; $I_{N}$ is the rated current of PMSG-WTG.

The PMSG-WTG adopts the reactive power priority control strategy during faults. The reference value of $d$-axis current is:
\begin{equation}
\begin{aligned}
I_{dref}=\min\{I_{dref1}\text{,}I_{dmax}\}\\
I_{dmax}=\sqrt{I_{max}^2-I_{qref}^2}\label{eq6}
\end{aligned}
\end{equation}
where $I_{dref1}$ is obtained by the constant dc-link voltage control, which is the same as the control strategy during normal operation; $I_{dmax}$ is the maximum value of $d$-axis current; $I_{max}$ is the maximum current allowed through the converter.

\begin{figure*}[t]
\centering
\includegraphics[width=7 in]{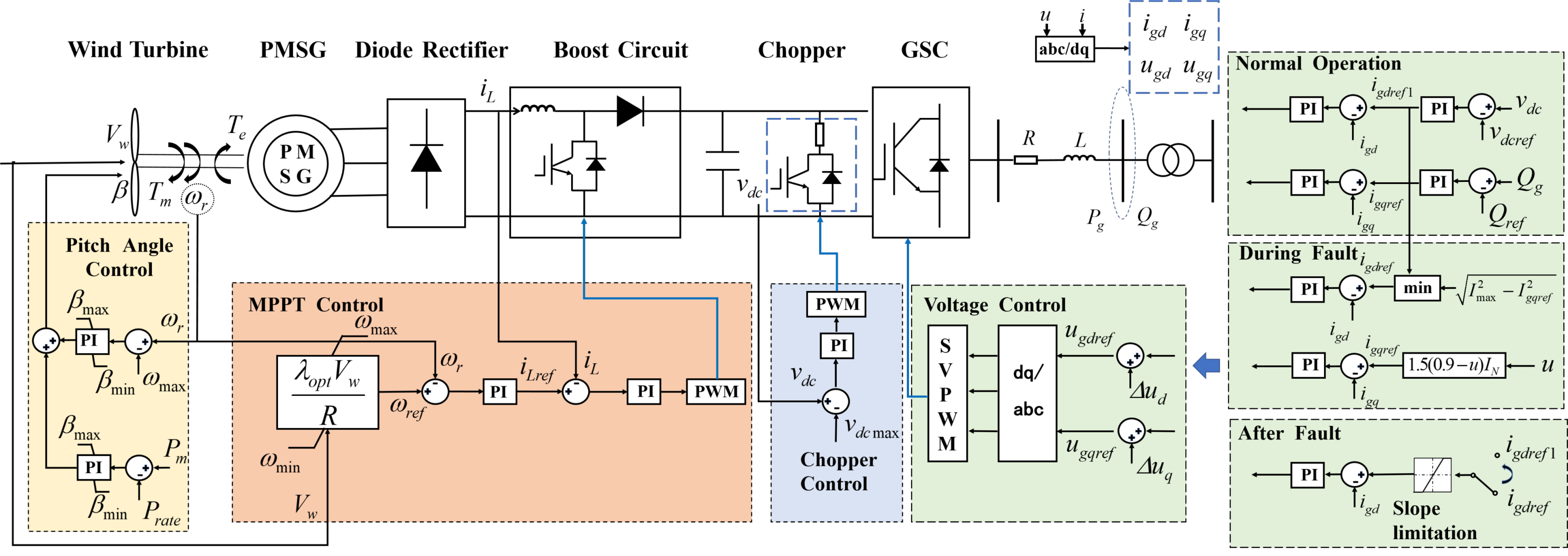}
\caption{Topology of the PMSG-WTG.}
\label{fig_2}
\end{figure*}

After the fault clearance, if the active power does not restore to its pre-fault value,   it will ramp up to the pre-fault value by limiting the recovering rate of $d$-axis current \cite{fortmann2013new,lorenzo-bonache_generic_2019,__2021}, which can reduce the mechanical stress \cite{feltes_comparison_2009}. The topology of the PMSG-WTG is shown in Fig. \ref{fig_2}

%

\section{ Classification and Discrimination Method of Active Power Dynamic Response Characteristics}
The active power dynamic response characteristics of the PMSG-WTG after a fault can be divided into two parts: the characteristics during the fault and the recovery characteristics after the fault is cleared, which will be analyzed respectively based on a single PMSG-WTG in the following parts.
\subsection{ Active Power Dynamic Response Characteristics During a Fault}
The output active power of GSC is instantaneously reduced to $\alpha P$ when the grid-side voltage drops to $\alpha$ due to an external fault. To maintain the dc voltage at the reference value, the steady-state value of $I_{dref1}$ is:
\begin{equation}
I_{dref1}=I_{d0}/ \alpha\label{eq9}
\end{equation}
where $I_{d0}$ is the pre-fault $d$-axis current of grid side.
Due to the fast response of PMSG-WTGs, it can be assumed that PMSG-WTGs are able to adjust the outputs $dq$-axis currents to the reference value \cite{guo2021data}. According to \eqref{eq6} and \eqref{eq9}, the reference value of $d$-axis current at the moment before fault clearance can be calculated by:

\begin{equation}
\begin{aligned}
& I_{dref1}^{f^-}=\frac{I_{d0}}{\alpha^{f^-}}\\
& I_{dref}^{f^-}=\min\{I_{dref1}^{f^-}\text{,}I_{dmax}^{f^-}\}
\end{aligned}
\label{add2}
\end{equation}
where the superscript $f^-$ denotes the moment before the fault clearance; $\alpha^{f^-}$ is the terminal voltage magnitude at the moment before the fault clearance; $I_{dref1}^{f^-}$ is the reference value obtained by the constant dc-link voltage control at the moment before the fault clearance; $I_{dref}^{f^-}$ and $I_{dmax}^{f^-}$ are the reference value and the allowed maximum value of the $d$-axis current of GSC at the moment before the fault clearance.

It can be known whether the output active power can recover to the pre-fault value during a fault by comparing $I_{dref1}^{f^-}$ and $I_{dmax}^{f^-}$. 


\subsection{ Active Power Dynamic Response Characteristics During Fault Recovery}
The main difference of the active transient response characteristics during fault recovery is the presence or absence of the active power ramp recovery process. Since the recovery rate of $d$-axis current is limited, the existence of active power ramp recovery process depends on whether the active power at the moment of fault clearance restores to the pre-fault value. That is, whether the $I_{dref}^{f^-}$ restores to $I_{d0}$. Since $I_{dref1}^{f^-}$ is always greater thatn $I_{d0}$, the existence of the ramp recovery process for the PMSG-WTG can be determined by comparing $I_{dmax}^{f^-}$ and $I_{d0}$.

 \subsection{Complete Active Power Dynamic Response Characteristics}\label{sec3.3}
Considering the active power dynamic response characteristics of PMSG-WTG during the fault duration and the fault recovery, the active power characteristics of PMSG-WTGs can be classified into the following three categories:
\begin{itemize}
	\item[1)] When $I_{dmax}^{f^-}<I_{d0}$, we can get $I_{dref}^{f^-}=I_{dmax}^{f^-}$. There is a ramp recovery process after the fault clearance.
	\item[2)] When $I_{d0}\le I_{dmax}^{f^-}<I_{dref1}^{f^-}$, we can get $I_{dref}^{f^-}=I_{dmax}^{f^-}$. The output active power of PMSG-WTG is lower than the pre-fault active power during the fault. And the output active power rises above the pre-fault value after the fault clearance and will return to the pre-fault value after a short period of oscillation.
	\item[3)] When $I_{dmax}^{f^-} \ge I_{dref1}^{f^-}$, we can get  $I_{dref}^{f^-}=I_{dref1}^{f^-}$. The output active power of PMSG-WTG has recovered to the pre-fault value during the fault, which can keep the dc-link voltage at the reference value. 
\end{itemize}

\begin{figure}[!htb]
	\centering
	\includegraphics[width=3.5 in]{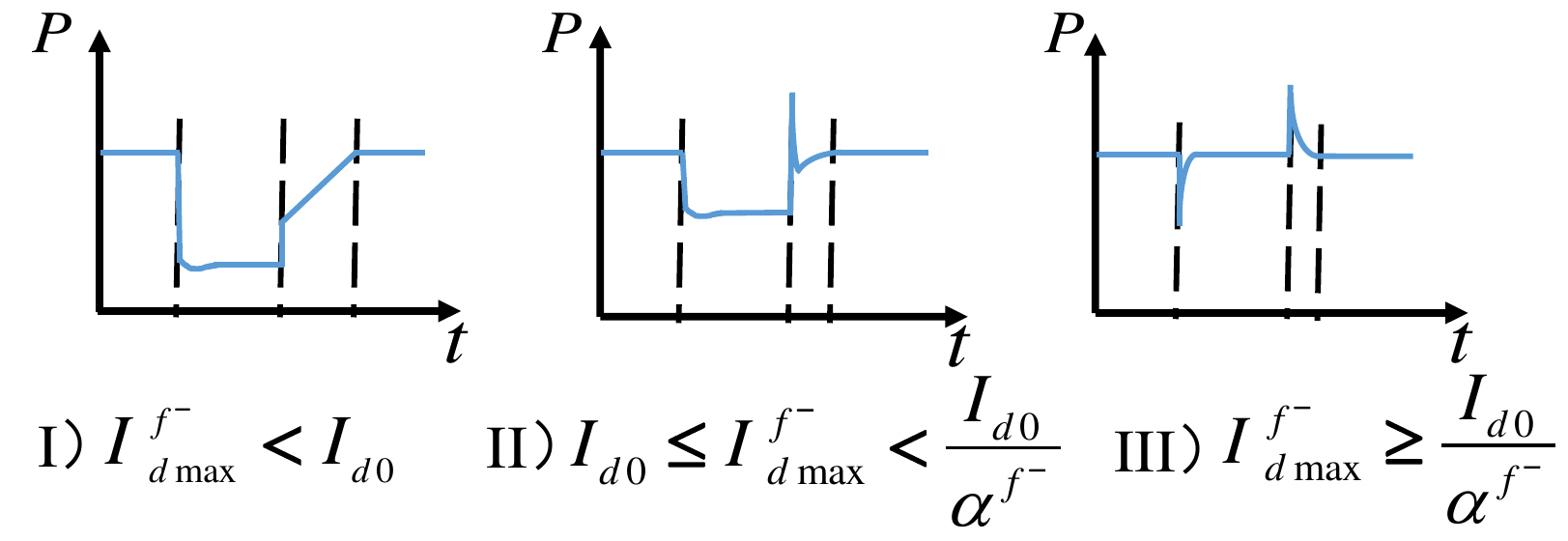}
	\caption{ Schematic diagram of the active power dynamic response curve of each cluster.}
	\label{fig_5}
\end{figure}

The schematic diagram of the active power dynamic response curve of each cluster is shown in Figure \ref{fig_5}.
 \subsection{Discrimination method of Active Power Dynamic Response Characteristics}
 During the normal operation, the $d$-axis current of PMSG-WTG can be derived by:
 \begin{equation}
 I_{d0}=\frac{2P_{0}}{3e}\label{eq10}
 \end{equation}
where $P_{0}$ is the pre-fault active power of PMSG-WTG.

According to \eqref{eq5} and \eqref{eq6}, $I_{dmax}^{f^-}$ can be derived by:
 \begin{equation}
I_{dmax}^{f^-}=\sqrt{I_{max}^2-2.25\times (0.9-\alpha^{f^-})^2I_{N}^2}\label{eq11}
\end{equation}

When $I_{dmax}^{f^-}=I_{d0}$, the critical pre-fault active power can be calculated by \eqref{eq10} and \eqref{eq11}:
 \begin{equation}
P_{cri1}=\frac{3}{2}e\sqrt{I_{max}^2-2.25\times (0.9-\alpha^{f^-})^2I_{N}^2} \label{eq12}
\end{equation}

When $I_{dmax}^{f^-}=I_{dref1}^{f^-} $, the second critical pre-fault active power is:
\begin{equation}
P_{cri2}=\alpha^{f^-} P_{cri1} \label{eq13}
\end{equation}

$I_{max}$ and $I_{N}$ can be known when the PMSG-WTG model is determined and the scale value of $e$ during normal operation is close to 1. The critical power can be calculated at different voltage drop degrees. And the critical power can be compared with $P_{0}$ to determine which category a PMSG-WTG belongs to. When $P_{0}>P_{cri1}$, the PMSG-WTG belongs to the category I in Figure. \ref{fig_5}; when $P_{cri2}<P_{0} \le P_{cri1}$, the PMSG-WTG belongs to the category II; and when $P_{0}<P_{cri1}$, the PMSG-WTG belongs to the category III.

Moreover, the critical wind speeds of each sub-group can be calculated from the critical power and the wind power curve of the PMSG-WTG:
\begin{equation}
\begin{aligned}
V_{cri1}=f^{-1}(P_{cri1})\\
V_{cri2}=f^{-1}(P_{cri2})\label{eq14}
\end{aligned}
\end{equation}

\begin{figure}[!htb]
	\centering
	\includegraphics[width=2.5 in]{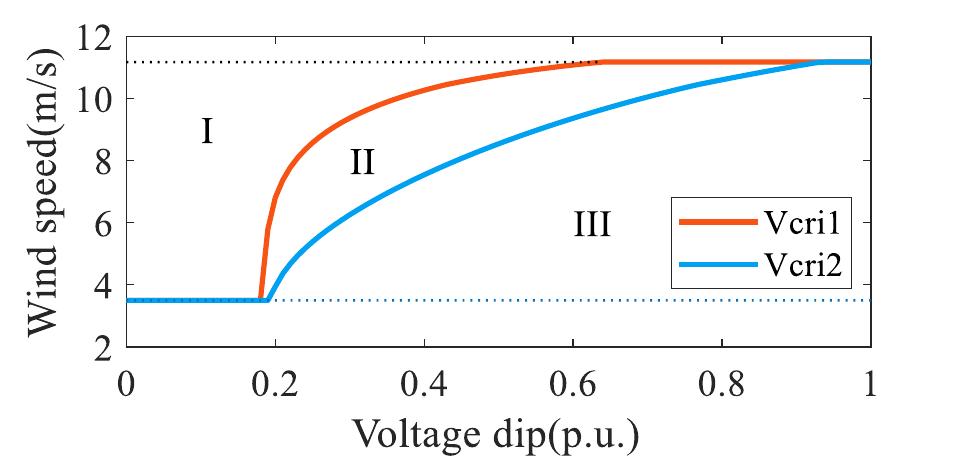}
	\caption{Classification boundaries of response characteristics.}
	\label{fig_7}
\end{figure}

According to \eqref{eq12} -\eqref{eq14}, the wind speed boundary conditions for each sub-group under different voltage dips  are shown in Fig. \ref{fig_7}.

The three regions in Figure. \ref{fig_7} correspond to the three types of response characteristics in Figure. \ref{fig_5}. The classification boundary here only shows the part of wind speeds from the cut-in wind speed of 3.5m/s to the rated wind speed of 11.1m/s. When the wind speed is below the cut-in wind speed, the output power of PMSG-WTG is 0. When the wind speed is above the rated wind speed, the output power of PMSG-WTG is the same as the active power of PMSG-WTG operating at the rated wind speed due to the pitch angle control.
\section{ Dynamic Equivalent Method for Wind Farm}
Based on the proposed clustering method in Section III, a single-machine equivalent method is designed in this section. In addition, for the PMSG-WTGs in the cluster I, a single-machine equivalent model with segmented ramp rate limitation for active current is introduced. Eventually, a collector line equivalent method is presented.
\subsection{ Single-Machine Equivalent Method}
\begin{figure}[!htb]
	\centering
	\includegraphics[width=2.5 in]{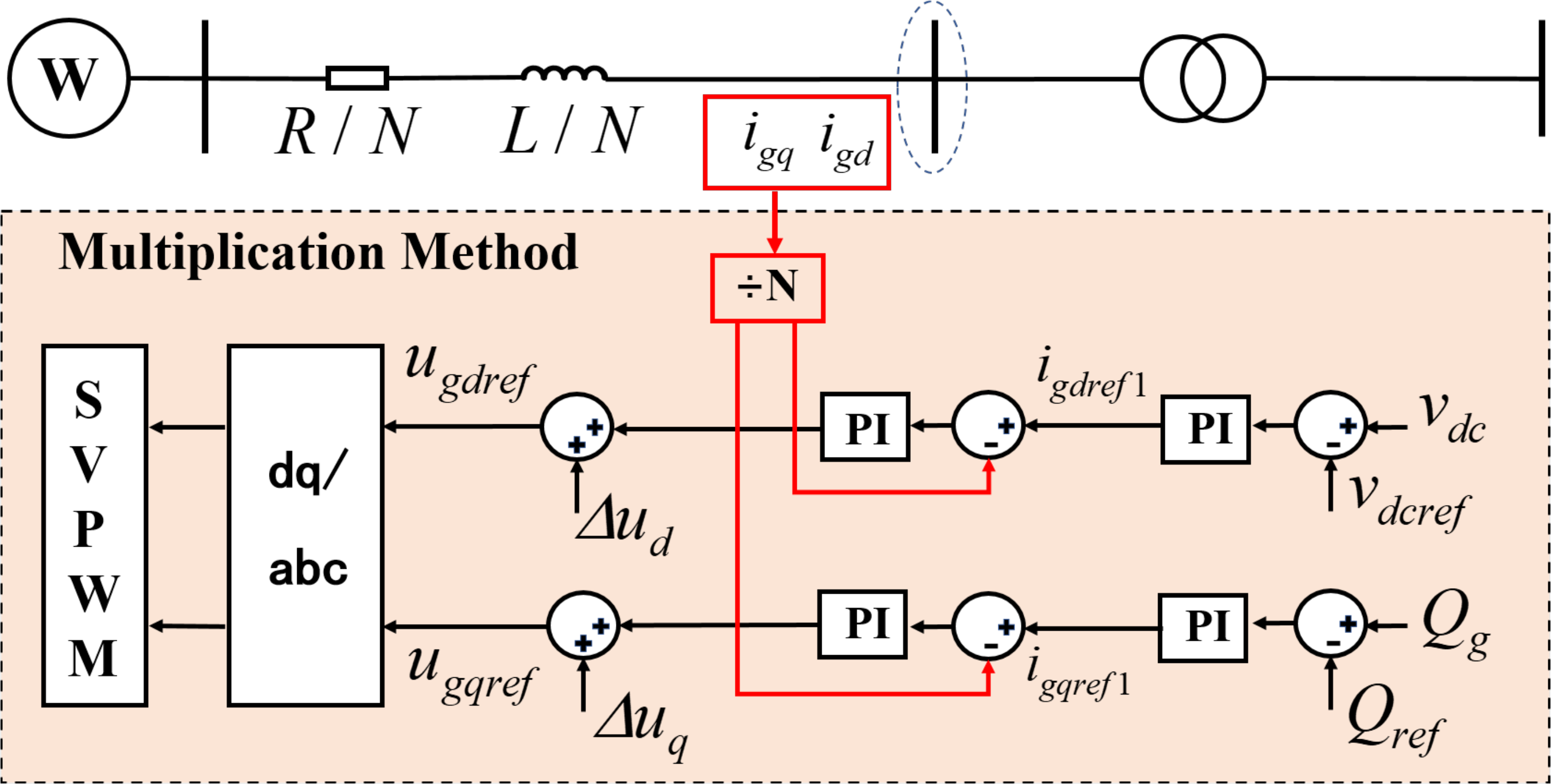}
	\caption{Single-machine equivalent method.}
	\label{fig_8}
\end{figure}

By dividing the measured $dq$-axis currents by the number of PMSG-WTGs ($N$) and inputting them to the control system, the equivalent model can multiply the output current of a single PMSG-WTG. And the stator-side impedance parameter of the equivalent model needs to be divided by $N$ to make the terminal voltage of equivalent model consistent with that of a single PMSG-WTG. In this case, the output power of the equivalent model is $N$ times the power of a single PMSG-WTG. The equivalent model is implemented as shown in the Fig. \ref{fig_8}.

In order to output the same active power in the steady state before and after the equivalence, the equivalent wind speed of the equivalent model should be derived by:
\begin{equation}
V_{eq,c}=f^{-1}(\frac{1}{N_c} \sum_{i=1}^{N_c}{f(V_{i,c})}) \ ,\  c=1,2,3\label{eq15}
\end{equation}
where $V_{eq,c}$ is the equivalent wind speed of cluster $c$; $N_c$ is the total number of PMSG-WTGs in cluster $c$; $i$ is the number of the PMSG-WTG; $V_{i,c}$ is the wind speed of the $i$th PMSG-WTG in cluster $c$.

For the PMSG-WTGs in cluster II and cluster III, their active power response characteristics are still basically the same as the active power of a single PMSG-WTG after accumulating. Therefore, the proposed single-machine equivalent method can be used to perform a proper equivalence for these two groups of PMSG-WTGs. However, for the PMSG-WTGs in cluster I, the time duration of their ramp recovery process varies due to the different initial operating wind speeds and voltage dips. By summing their active power during recovery process, the equivalent response curve will restore to pre-fault value with multiple different rates, which cannot be represented in the single-machine equivalent model mentioned above. As a result, there is still a need to further propose an equivalent method for the PMSG-WTGs in cluster I that can accurately represent the response characteristics during the fault recovery.
\subsection{Single-machine Equivalent method with Multi-segment Slope Limitation}
\begin{figure}[!htb]
	\centering
	\includegraphics[width=3 in]{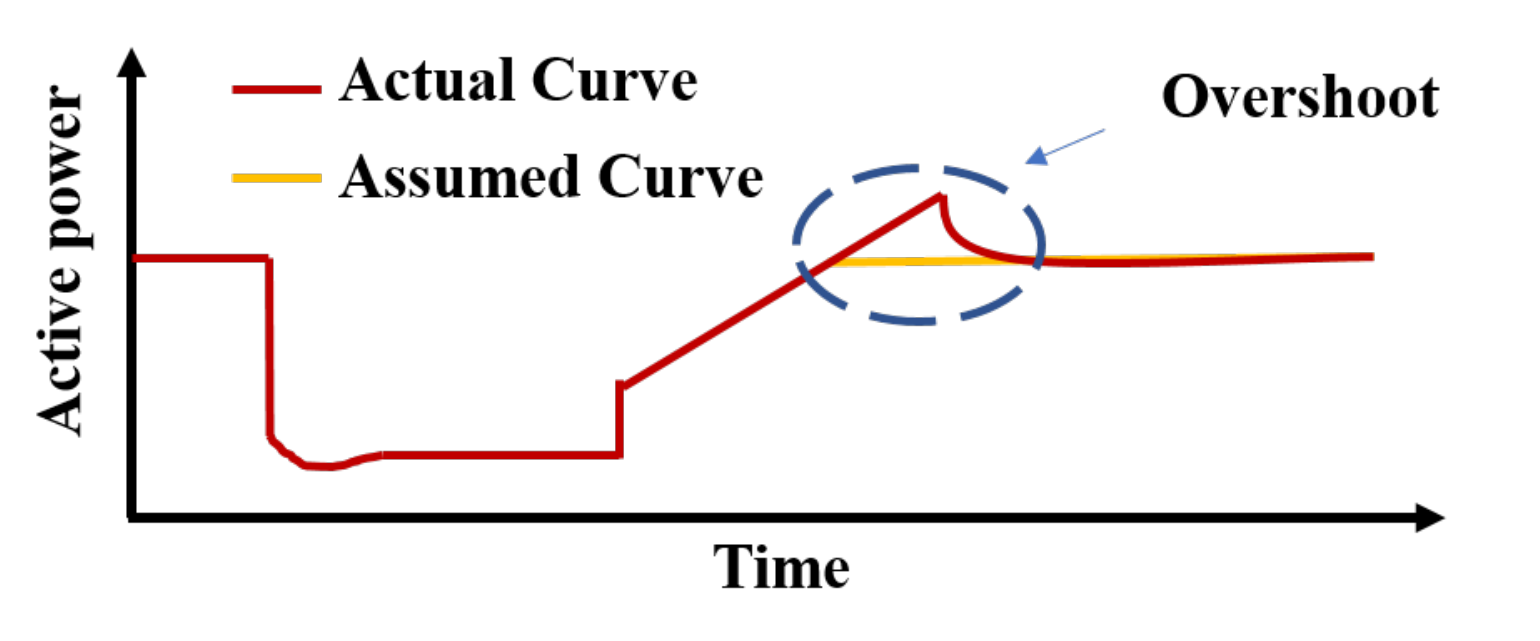}
	\caption{Schematic diagram of the assumed curve.}
	\label{fig_9}
\end{figure}

 In order to reflect the differences in recovery characteristics of each PMSG-WTG in cluster I, a single-machine equivalent method with multi-segment slope limitation is proposed. Ignoring the overshoot of active power during constant dc voltage control, it can be assumed that the ramp recovery process ends when the active power rises to the pre-fault value as shown in Fig. \ref{fig_9}. And the calculation method for the time duration of the ramp recovery process is introduced below.

The pre-fault active power and pre-fault $d$-axis current of the $i$th PMSG-WTG in the cluster I can be derived as follows:
\begin{equation}
\begin{aligned}
P_{0i}=f(V_{wi}) \\
I_{d0i}=\frac{2P_{0i}}{3e} \label{eq16}
\end{aligned}
\end{equation}

The $q$-axis current to be supplied by the $i$th PMSG-WTG during a fault is:
\begin{equation}
I_{qi}=1.5(0.9-\alpha ^{f^-}_i)I_N,(0.2\le \alpha ^{f^-}_i \le 0.9) \label{eq17}
\end{equation}
where $\alpha ^{f^-}_i$ and $I_{qi}$ are the terminal voltage magnitude and the reference value of $q$-axis current of the $i$th PMSG-WTG at the moment before the  fault clearance. 

Since the $d$-axis  of the PMSG-WTGs in cluster I all reach $I_{dmax}^{f^-}$ during the fault, the $d$-axis current of the $i$th PMSG-WTG can be derived as follows according to \eqref{eq11}:
\begin{equation}
I_{di}^{f^-}=\sqrt{I_{max}^2-2.25\times (0.9-\alpha ^{f^-}_i)^2I_{N}^2}\label{eq18}
\end{equation}

Due to the limitation of the recovery rate of the $d$-axis current, the time duration of the ramp recovery process of the $i$th PMSG-WTG is:
\begin{equation}
t_i=(I_{d0i}-I_{di}^{f^-})/k \label{eq19}
\end{equation}
where $k$ is the maximum recovery rate of $d$-axis current; $t_i$ is the time duration of the slope recovery process of the $i$th PMSG-WTG.

Substituting \eqref{eq16} and \eqref{eq18} into \eqref{eq19}, the time duration of the $i$th PMSG-WTG can be derived as follows:
\begin{equation}
t_i=(\frac{2f(V_{wi})}{3e}-\sqrt{I_{max}^2-2.25\times (0.9-\alpha ^{f^-}_i)^2I_{N}^2})/k \label{eq20}
\end{equation}

\begin{figure}[!htb]
	\centering
	\includegraphics[width=3 in]{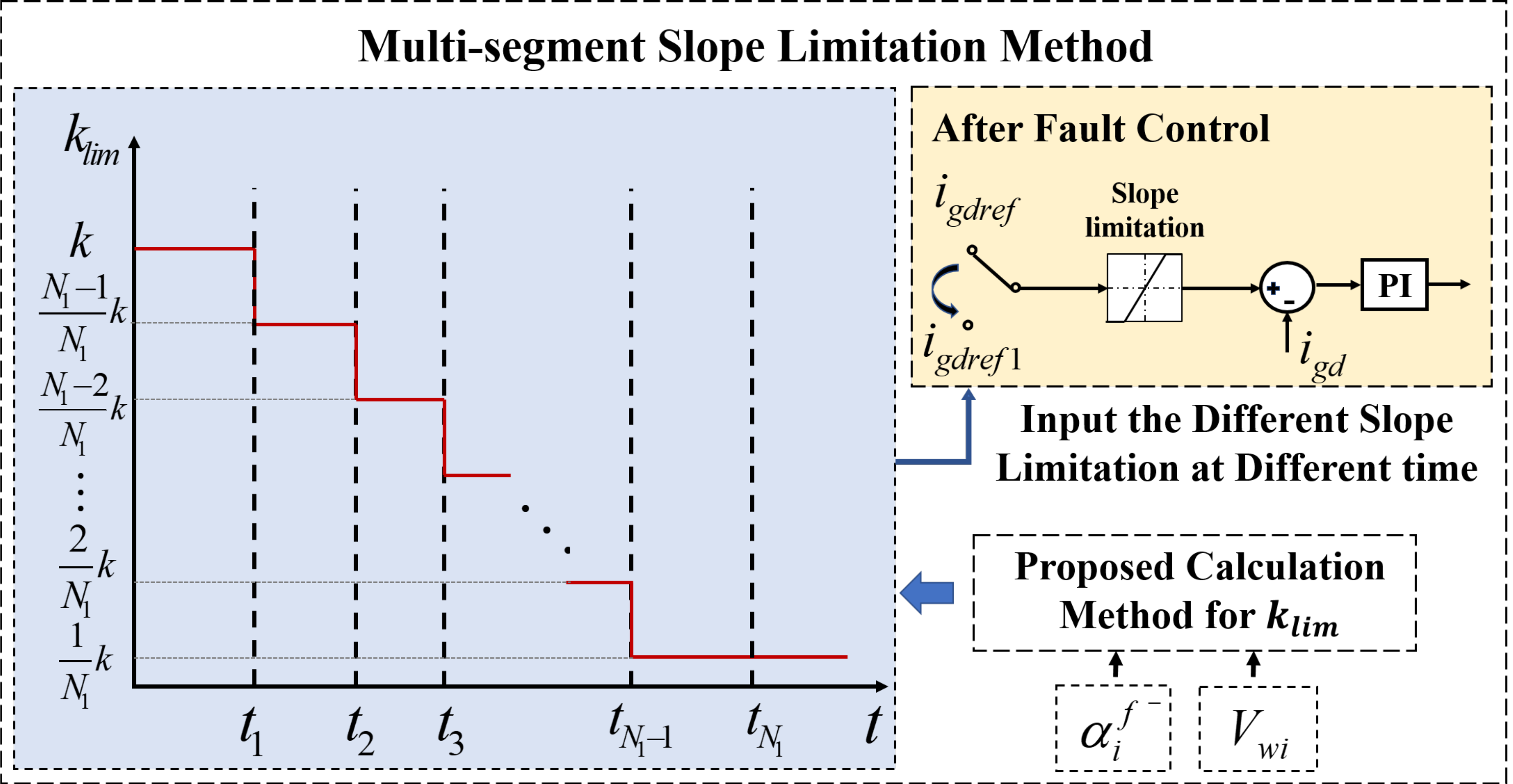}
	\caption{Multi-segment slope limitation method.}
	\label{fig_10}
\end{figure}

According to \eqref{eq20}, $t_i$ is only related to $\alpha ^{f^-}_i$ and $V_{wi}$. Sorting $t_i$ from the smallest to the largest and the recovery rate limitation of $d$-axis current of the equivalent model after fault clearance is as follows:
\begin{equation}
k_{lim}=
\begin{cases}
\ \ \ \ \ \ \  k & ,  t <t_1\\
(N_1-j)k/N_1 & , t_j\le t<t_{j+1},j=1\cdots N_1-1\\
\ \ \ \ \ \ k/N_1 & , t\ge t_{N_1}
\label{eq21}
\end{cases}
\end{equation}
where $N_1$ is the number of PMSG-WTGs in cluster I. And the illustration of the proposed method is shown in Fig. \ref{fig_10}.

\subsection{Equivalent Method for Collector Lines}
\begin{figure}[htb!]
	\centering
	\includegraphics[width=3 in]{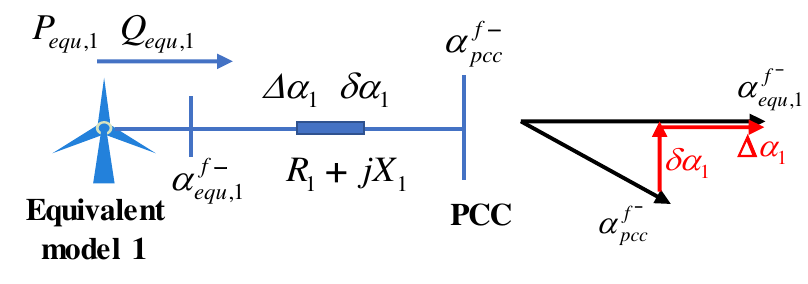}
	\caption{Equivalent model and collector line of cluster I}
	\label{collector_line}
\end{figure}
When an external fault occurs, the output reactive power of the $i$th PMSG-WTG is:
\begin{equation}
Q_i=1.5I_{qi}\alpha ^{f^-}_i \label{eq22}
\end{equation}

The output reactive power of the equivalent model can be obtained by summing up the reactive power of each PMSG-WTG, as presented by:
\begin{equation}
Q_{equ,c}=\sum_{i=1}^{N_c}Q_i \ ,\  c=1,2,3\label{eq23}
\end{equation}

 The schematic diagram of the equivalent model and the collector line of cluster I is shown in the Fig. \ref{collector_line}. In order to keep the output reactive power the same before and after the equivalence, the terminal voltage drop of the equivalent model should satisfy the following equation:
\begin{equation}
Q_{equ,c}=1.5\times1.5(0.9-\alpha ^{f^-}_{equ,c})I_N\alpha ^{f^-}_{equ,c}N_c \label{eq24}
\end{equation}

$\alpha ^{f^-}_{equ,c}$ is solved as:
\begin{equation}
\alpha ^{f^-}_{equ,c\ 1,2}=\frac{0.9\pm\sqrt{0.9^2-4\frac{Q_{equ,c}}{2.25I_NN_c}}}{2} \label{eq25}
\end{equation}
where $\alpha ^{f^-}_{equ,c\ 1,2}$ are the two solutions of \eqref{eq24}. And $\alpha ^{f^-}_{equ,c}$ is finally determined by:
\begin{equation}
\alpha ^{f^-}_{equ,c}=
\begin{cases}
\alpha ^{f^-}_{equ,c\ 1} & , \lvert \alpha ^{f^-}_{equ,c\ 1}-\alpha ^{f^-}_{pcc}\rvert \le \lvert \alpha ^{f^-}_{equ,c\ 2}-\alpha ^{f^-}_{pcc}\rvert \\
\alpha ^{f^-}_{equ,c\ 2} & , \lvert \alpha ^{f^-}_{equ,c\ 1}-\alpha ^{f^-}_{pcc}\rvert > \lvert \alpha ^{f^-}_{equ,c\ 2}-\alpha ^{f^-}_{pcc}\rvert
\label{eq26}
\end{cases}
\end{equation}
where $\alpha ^{f^-}_{pcc}$ is the voltage of PCC; $\alpha ^{f^-}_{equ,c}$ is the terminal voltage of the equivalent model of cluster $c$.

From the conclusion of Section III.C, the active power of equivalent model of each cluster at the moment of fault clearance can be expressed by:
\begin{equation}
\begin{cases}
P_{equ,1}=1.5N_1\alpha ^{f^-}_{equ,1}\sqrt{I_{max}^2-2.25\times (0.9-\alpha ^{f^-}_{equ,1})^2I_{N}^2} \\
P_{equ,2}=1.5N_2\alpha ^{f^-}_{equ,2}\sqrt{I_{max}^2-2.25\times (0.9-\alpha ^{f^-}_{equ,2})^2I_{N}^2} \\
P_{equ,3}=N_3f(V_{eq,3})
\label{eq27}
\end{cases}
\end{equation}
where $P_{equ,c}$ is the output active power of equivalent model of cluster $c$; $\alpha ^{f^-}_{equ,1}$ and $\alpha ^{f^-}_{equ,2}$ are the terminal voltages of the equivalent models of cluster I and cluster II, respectively. The longitudinal component and the lateral component of the voltage drop on the equivalent collector line are:
\begin{equation}
\begin{cases}
\Delta \alpha_c=(P_{equ,c}R_c+Q_{equ,c}X_c)/\alpha ^{f^-}_{equ,c} \\
\delta \alpha_c=(P_{equ,c}X_c-Q_{equ,c}R_c)/\alpha ^{f^-}_{equ,c}
\label{eq28}
\end{cases}
\end{equation}
where $\Delta \alpha_c$ and $\delta \alpha_c$ are the longitudinal component and the lateral component of the voltage drop on the collector line of cluster $c$, respectively; $R_c$ and $X_c$ are the resistance and reactance of the equivalent line of cluster $c$.

The parameters of collector line are designed to make the terminal voltage drop of the equivalent model equal to $\alpha ^{f^-}_{equ,c}$. Thus, the parameters of collector line will satisfy the following equation: 
\begin{equation}
(\alpha ^{f^-}_{equ,c}-\Delta \alpha_c)^2+\delta \alpha_c^2=\alpha ^{f^-2}_{pcc}
\label{eq29}
\end{equation}

The \eqref{eq28} and \eqref{eq29} has three equations with four unknowns ($\Delta \alpha_c$, $\delta \alpha_c$, $R_c$ and $X_c$), which means an additional equation is still required. In this article, the per unit length impedance parameters of the equivalent collector line are designed the same as that before the equivalence. And only the line length is changed to satisfy \eqref{eq29}. That is to say, the additional equation is:
\begin{equation}
\frac{R_c}{X_c}=\frac{R_0}{X_0}=K_0
\label{eq30}
\end{equation}
where $R_0$ and $X_0$ are the per unit length resistance and reactance of collector line before the equivalence.

\begin{figure}[!htb]
	\centering
\vspace{-0.3cm}  
	\includegraphics[width=2 in]{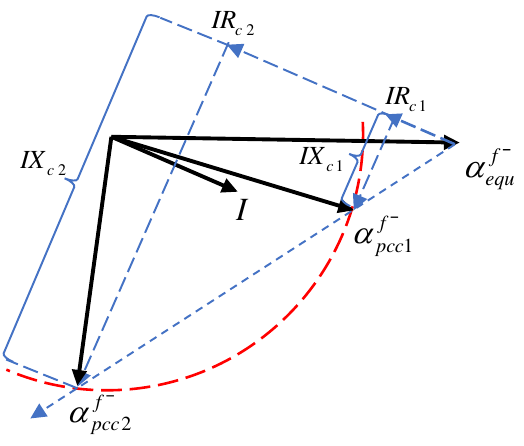}
	\caption{Vector diagram of the solutions}
	\label{fig_11}
\end{figure}

According to \eqref{eq28}-\eqref{eq30}, the parameters of equivalent line of each cluster are:
 \begin{equation}
 \begin{aligned}
 & X_{c\ 1,2}=\frac{\alpha ^{f^-}_{equ,c}(A\mp \sqrt{A^2-B(\alpha ^{f^-2}_{equ,c}-\alpha ^{f^-2}_{pcc})})}{B}\\
 & R_{c\ 1,2}=K_0X_{c\ 1,2}\\
 & A=\alpha ^{f^-}_{equ,c}(K_0P_{equ,c}+Q_{equ,c})\\
 & B=(P_{equ,c}^2+Q_{equ,c}^2)(1+K_0^2)
 \label{eq31}
 \end{aligned}
 \end{equation}
 where $X_{c\ 1,2}$ and $R_{c\ 1,2}$ are the two sets of solutions of equations, as shown in Fig. \ref{fig_11}. The phase angle difference between $\alpha ^{f^-}_{equ,c}$ and $\alpha _{pcc2}^{f^-}$ exceeds $\pi/2$, which does not meet the requirements of static stability \cite{araposthatis_analysis_1981}. Thus, $R_{c\ 1}$ and $X_{c\ 1}$ are chosen as the equivalent line parameters, that is:
  \begin{equation}
 \begin{aligned}
 & X_{c}=\frac{\alpha ^{f^-}_{equ,c}(A-\sqrt{A^2-B(\alpha ^{f^-2}_{equ,c}-\alpha ^{f^-2}_{pcc})})}{B}\\
 & R_{c}=K_0X_{c}\\
 \label{result_XR}
 \end{aligned}
 \end{equation}
\vspace{-0.3cm}
\section{Calculation Method for Terminal Voltages of PMSG-WTGs}
Wind speeds and terminal voltages are the clustering indicators of the dynamic equivalent method proposed in Section III and Section IV. When simulation analysis is performed for  expected contingencies, the operating wind speed of each PMSG-WTG can be obtained by the historical wind speed data or wind speed forecast \cite{lei2009review,zolfaghari2015new}. However, when an external fault occurs in a real power system, terminal voltages of PMSG-WTGSs at the moment of fault clearance cannot be predicted in advance because it is not only related to the output characteristics of PMSG-WTG, but also to the time duration of the fault and the response characteristics of the components on the system side. Thus, it is difficult to solve the voltages analytically. Therefore, this section proposes a method for solving the terminal voltage of each PMSG-WTG in a wind farm when the PCC voltage is given. And a simulation-based iterative method is proposed to get the value of PCC voltage.
\subsection{A Calculation Method for PMSG-WTG Terminal Voltages}
\begin{figure}[!htb]
	\centering
	\includegraphics[width=3 in]{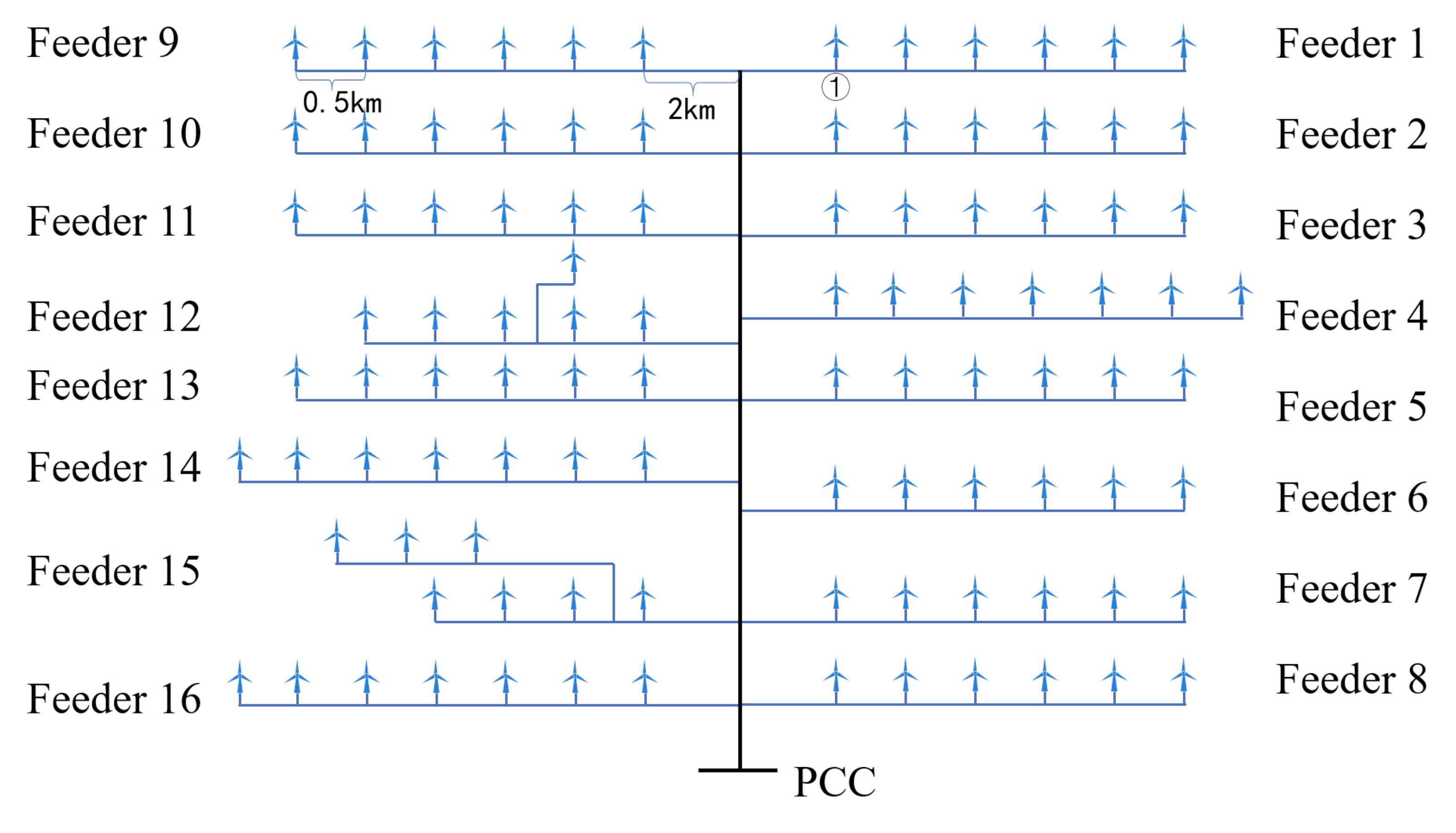}
	\caption{Topology of wind farm.}
	\label{fig_12}
\end{figure}

This section presents a calculation method for the terminal voltage of each PMSG-WTG based on a real wind farm topology in China. The topology of the wind farm is shown in Fig. \ref{fig_12}.

When the voltage of PCC is given, the calculation of the terminal voltages at a certain moment is similar to the power flow calculation assuming that the PMSG-WTGs have adjusted the output $dq$-axis currents to the reference value. 

According to \eqref{eq3}, \eqref{eq5} and \eqref{add2}, the output active and reactive power of each PMSG-WTG are:
\begin{equation}
\begin{aligned}
& P_i=1.5\lvert \dot U_i \rvert \times\min\{\frac{I_{d0i}}{\lvert \dot U_i \rvert}\text{,}I_{dmaxi}\} \\
& Q_i=1.5\lvert \dot U_i \rvert I_{qrefi} \\
& I_{qrefi}=1.5\times (0.9-\lvert \dot U_i \rvert)I_{N},(0.2\le \lvert \dot U_i \rvert \le 0.9)\\
& I_{dmaxi}=\sqrt{I_{max}^2-I_{qrefi}^2}
\label{eq32}
\end{aligned}
\end{equation}
where $\dot U_i $ is the voltage phasor of terminal voltage of the $i$th PMSG-WTG; $P_i$ and $Q_i$ are the  output active and reactive power of the $i$th PMSG-WTG, respectively. Then, the current injected by the $i$th PMSG-WTG is:
\begin{equation}
\dot I_{ni}=(\frac{P_i+j\Delta Q_i}{\dot U_i})^*
\label{eq33}
\end{equation}

%
%
where $\dot I_{ni}$ is the current injected by the $i$th PMSG-WTG.

To further calculate the current of each branch on the feeder, a branch-node matrix needs to be created. The rows of the matrix represent branches, and the columns of the matrix represent nodes. If the injected current of node $j$ flows through the $i$th branch, the value of the $i$-th row and $j$-th column of the matrix is 1, otherwise it is 0. 

The branch current column vector of a feeder is:
\begin{equation}
\dot I_{b}=C\dot I_{n}
\label{eq35}
\end{equation}
where $C$ is the branch-node matrix of a feeder line; $\dot I_n$ is column vector of the nodal injection current consisting of $\dot I_{ni}$; $\dot I_{b}$ is the branch current column vector. Further, based on the branch current, the branch voltage drop and node voltage drop are:
\begin{equation}
\begin{aligned}
& \Delta \dot U_b=Z \dot I_b \\
& \Delta \dot U_n=C^T\Delta \dot U_b \label{eq36}
\end{aligned}
\end{equation}
where $\Delta \dot U_b$ and $\Delta \dot U_n$ are the branch voltage drop and node voltage drop column vector of the feeder, respectively; $Z$ is the branch impedance matrix of feeder. Only the values of the diagonal elements of $Z$ are the branch impedance, and the values of other non-diagonal elements are 0. Then, the voltage of each node on the feeder can be obtained as:
\begin{equation}
\dot U^{'}=\dot U_{pcc}+\Delta \dot U_{n}
\label{eq37}
\end{equation}
where $\dot U_{pcc}$ is the voltage of PCC; $\dot U^{'}$ is the revised voltage of each node on the feeder.

Substituting \eqref{eq35} - \eqref{eq36} into \eqref{eq37}, the revised voltage can be derived as:
\begin{equation}
\dot U^{'}=\dot U_{pcc}+C^TZC\dot I_{n}
\label{eq38}
\end{equation}

When the voltage at PCC is known, the terminal voltage of each PMSG-WTG can be calculated by the following steps:
\begin{itemize}
	\item[1)] Branch impedance matrix and branch-node matrix are formed for each feeder based on the wind farm topology data. And set all terminal voltages of PMSG-WTGs to the PCC voltage. 
	\item[2)] Use equations \eqref{eq32}, \eqref{eq33} and \eqref{eq38} to calculate the revised terminal voltage of each PMSG-WTG.
	\item[3)] If $\lvert \dot U^{'}-\dot U \rvert<\sigma$, turn to step 4. If not, assign the value of $\dot U^{'}$ to $\dot U$ and turn to step 2. Where $\sigma$ is the iteration tolerance.
	\item[4)] Output the $\dot U^{'}$, and $\dot U^{'}_i$ is the terminal voltage of the $i$th PMSG-WTG.
\end{itemize}
\subsection{Simulation-based Iterative Method for Solving the Voltage at PCC}
 
As the PCC voltage is difficult to solve analytically, this section proposes a simulation-based iterative method for solving the voltage at PCC, which can improve both the computational efficiency and accuracy. The steps are as follows:
\begin{itemize}
	\item[1)] Input the wind speed of each PMSG-WTG during normal operation and let $\alpha ^{f^-}_{pcc}=1$.
	\item[2)] Let $\dot U_{pcc}=\alpha ^{f^-}_{pcc}\angle 0 ^\circ$, and calculate the terminal voltage of each PMSG-WTG using the method proposed in Section V.A.
	\item[3)] The equivalent model can be performed by the method proposed in Section III and Section IV.
	\item[4)] Simulation analysis of the expected contingency is carried out using the established equivalent model in step (3). And the PCC voltage at the moment of fault clearance ($\alpha ^{f^{-'}}_{pcc}$) can be obtained by the result of the simulation.
	\item[5)] If $\lvert \alpha ^{f^{-'}}_{pcc}-\alpha ^{f^-}_{pcc} \rvert<\sigma$, turn to step 6. If not, assign the value of $\alpha ^{f^{-'}}_{pcc}$ to $\alpha ^{f^-}_{pcc}$ and turn to step 2.
	\item[6)] Since the cluster indicators are known, the equivalent model can be obtained by the method proposed in Section III and Section IV.
\end{itemize}

\begin{figure}[!htb]
	\centering
	\includegraphics[width=3.5 in]{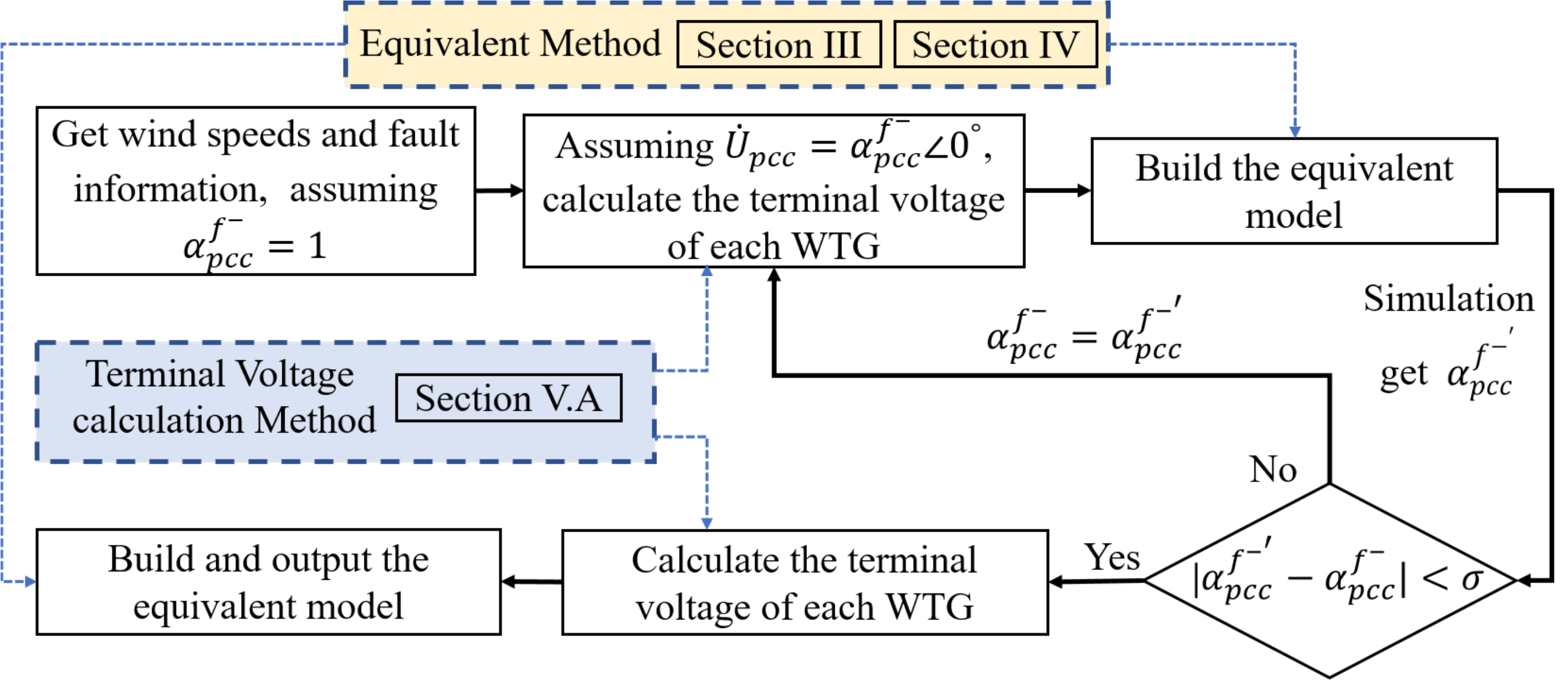}
	\caption{Flow chart of the proposed method.}
	\label{fig_14}
\end{figure}

The above method avoids the analytical solution for $\alpha ^{f^{-}}_{pcc}$. And in the case study section, it is demonstrated that the method can meet the convergence accuracy after 1-2 iterations. The overall flow chart of the proposed equivalent method is shown in the Fig. \ref{fig_14}. 

The branch-node matrices, branch impedance matrices and clustering boundaries can be get offline. Thus, the wind farm equivalent model can be rapidly obtained if the wind speeds and the fault information can be known. In general, the wind speeds can be obtained from wind speed forecasts and the fault information is known in the DSA. Therefore, the proposed equivalent method is adapted to the online DSA.
\section{Method Verification}
\begin{figure}[!htb]
	\centering

	\includegraphics[width=2.8 in]{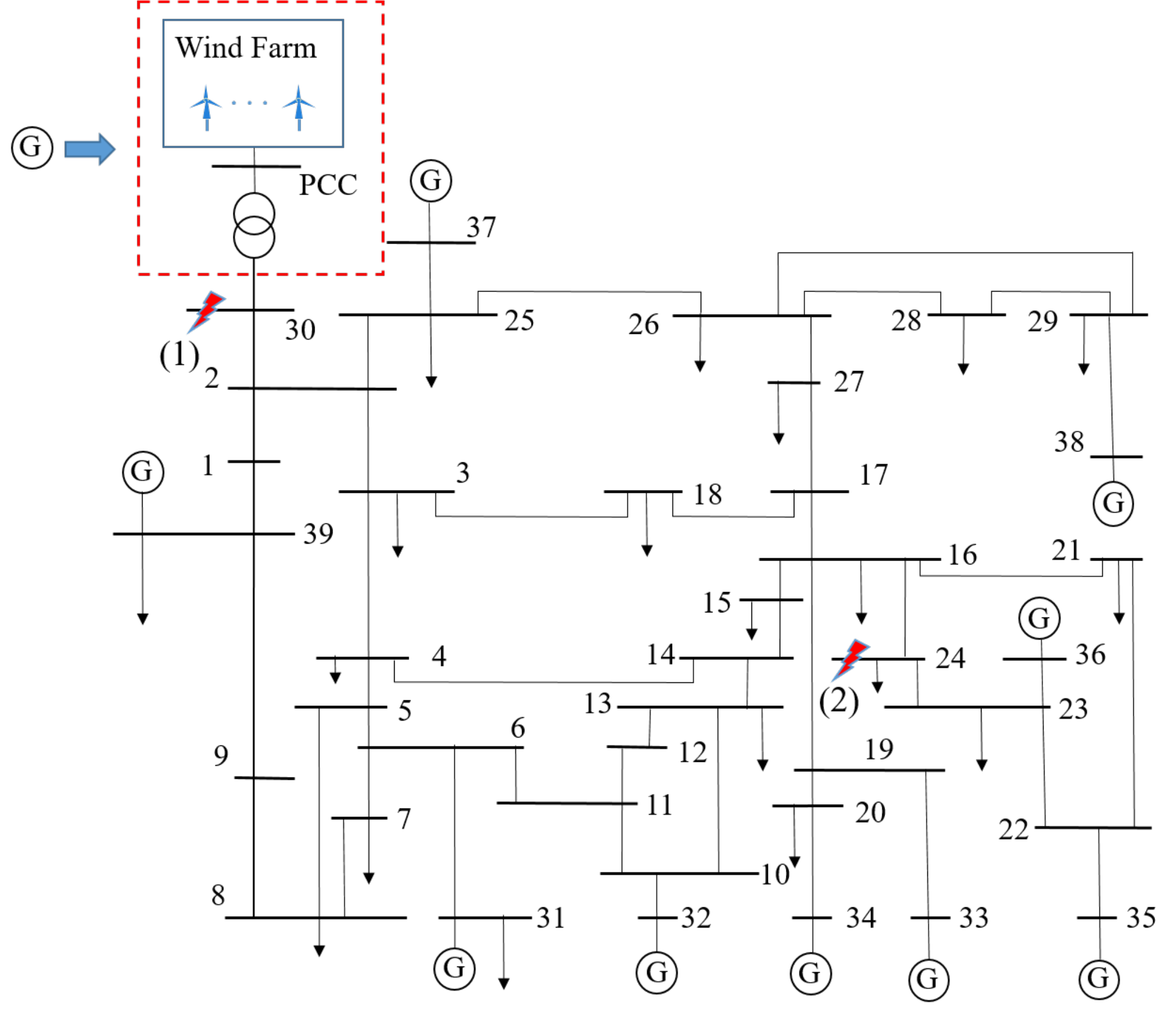}
	\caption{IEEE 39-bus system with a wind farm integrated.}
	\label{fig_15}
\end{figure}

In order to verify the proposed equivalent method, a study is conducted on an actual wind farm in China, as shown in Figure \ref{fig_12}. The wind farm includes 100 PMSG-WTGs with the rated capacity of 1.5MW. The distance between two wind turbines on the same feeder is 0.5km, and the distance between the feeder and the PCC is 2km. The detailed model of the wind farm is connected to node 30 of the IEEE 39-bus system through a transformer, replacing the original generator, as shown in Fig. \ref{fig_15}. 

\begin{figure}[!htb]
	\centering
\vspace{-0.2cm}  
\setlength{\belowcaptionskip}{-0.2cm}   
	\includegraphics[width=3 in]{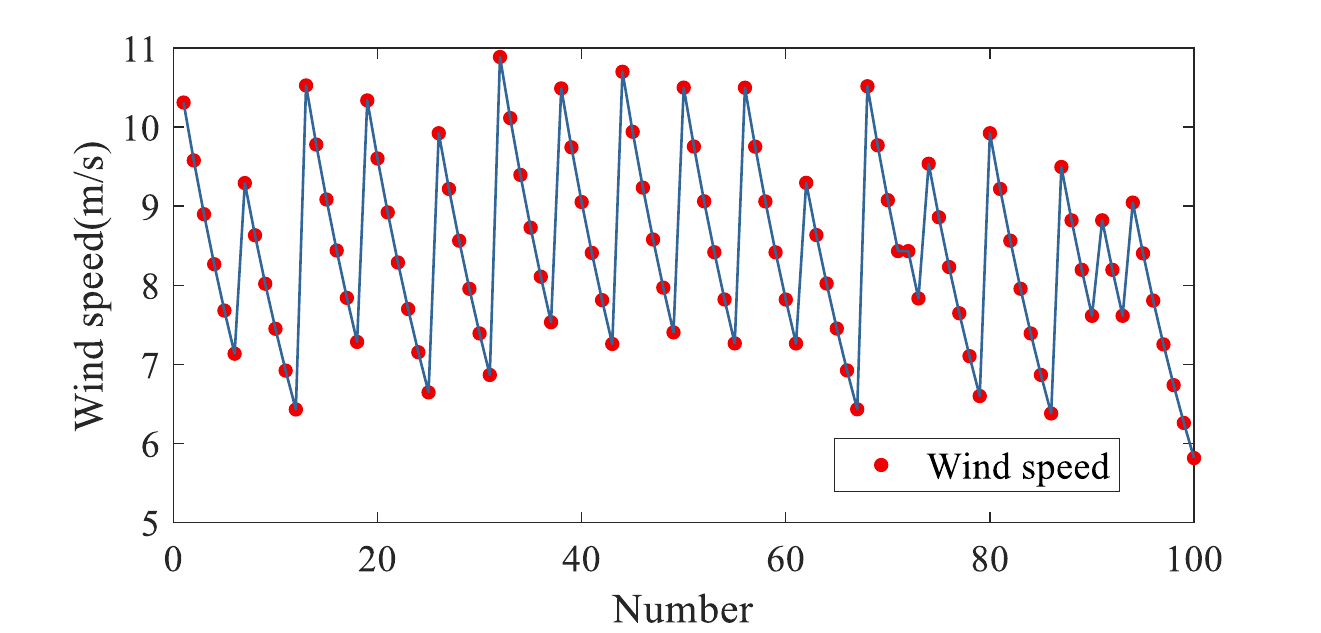}
	\caption{Wind speed of each PMSG-WTG.}
	\label{fig_16}
\end{figure}

Assuming that the distance between the feeders is far enough, wind speeds among different feeders do not affect each other. And the wind speeds of PMSG-WTGs on the same feeder are modeled with Jensen model, which can be derived as: 
\begin{equation}
\begin{aligned}
& V_{wi}=V_{w0}dec^{i-1}\\
& dec=(1-(1-(1-C_t)^\frac{1}{2})(\frac{\gamma}{\gamma+kx})^2)
\label{eq39}
\end{aligned}
\end{equation}
where $V_{w0}$ is the initial input wind speed of the feeder; $C_t$ is the thrust coefficient; $k$ is the wake decay constant; $\gamma$ is the radius of wind turbine.
The initial input wind speed of each feeder is an uniform random distribution between 9 and 11 m/s. The thrust coefficient is 0.2, and the wake decay constant is 0.04. The wind speeds of PMSG-WTGs using in the following case are shown in the Fig. \ref{fig_16}. The numbering of PMSG-WTGs starts at the node nearest to the PCC of feeder 1 and ends at the farthest node of feeder 16. The proposed equivalent method is validated below based on two different expected contingencies.
\subsection{Case I: Verification of the Method When a Nearby Fault Occurs}
The three-phase short circuit fault starts at 3.0s and clears at 3.1s at node 30. The PCC voltage is obtained by the method proposed in Section V.B. And the results of each iteration are shown in table. \ref{Voltage_iter}. 

\begin{table}
	\renewcommand\arraystretch{1.5}
	\begin{center}
		\caption{PCC Voltage in Each Iteration of Case I.}
		\label{Voltage_iter}
				\resizebox{0.99\linewidth}{0.06\textheight}{
		\begin{tabular}{| c | c | }
			\hline
			Model  & Simulation result of $\alpha ^{f^{-'}}_{pcc}$ (p.u.)\\
			
			\hline
			Initial equivalent model ($\alpha ^{f^{-'}}_{pcc}=1$) & 0.2238 \\
			\hline
			Equivalent model 1 ($\alpha ^{f^{-'}}_{pcc}=0.2238$) & 0.2250 \\ 
			\hline
			Equivalent model 2 ($\alpha ^{f^{-'}}_{pcc}=0.2250$) & 0.2250 \\
			\hline 
			Detailed model&0.2250\\ 
			\hline			
		\end{tabular}
	}
	\end{center}
\end{table}

\begin{figure}[!htb]
	\centering
	\includegraphics[width=3 in]{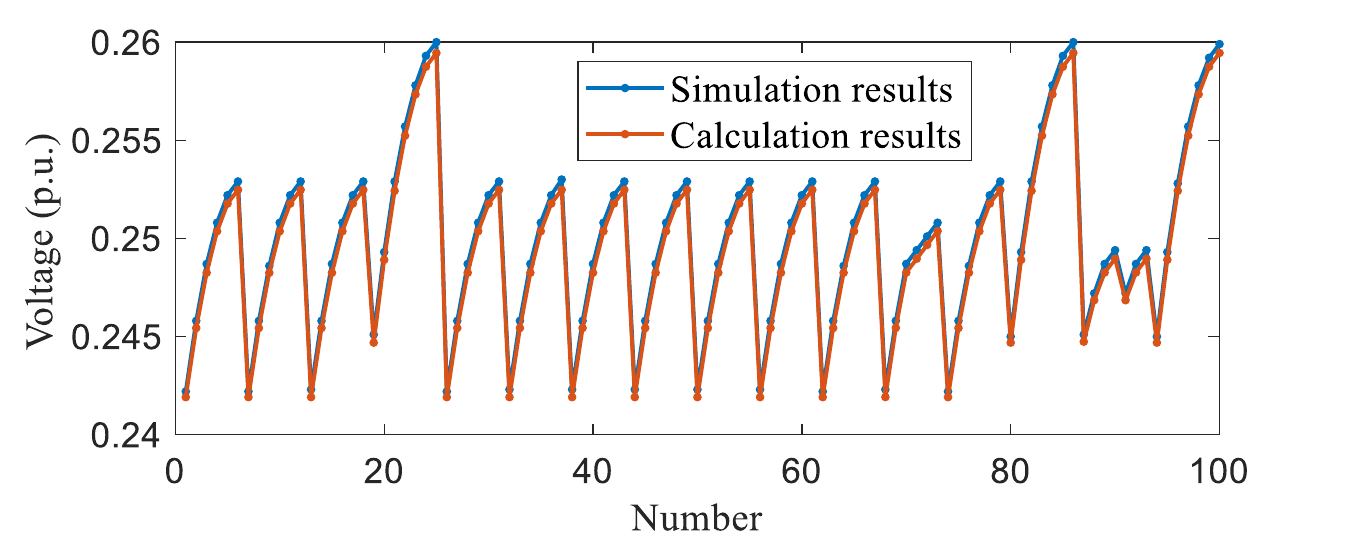}
	\caption{Results of terminal voltages.}
	\label{fig_17}
\end{figure}

When $\alpha ^{f^{-'}}_{pcc}=0.2250$, the terminal voltages of PMSG-WTGs at the moment before fault clearance can be calculated using the method proposed in Section V.A. The calculation results are compared with the simulation results of detailed model, as shown in figure. \ref{fig_17}. The maximum error percentage of all nodes is $0.21\%$, which proves the validity of the method proposed in Section V.A.

Further, the PMSG-WTGs are clustered based on the terminal voltages and wind speeds. The clustering results are shown in the table \ref{Clustering Results of Case I.}.
\begin{table}
	\renewcommand\arraystretch{1.2}
	\begin{center}
		\caption{Clustering Results of Case I.}
		\label{Clustering Results of Case I.}
				\resizebox{0.99\linewidth}{0.08\textheight}{
		\begin{tabular}{| c | c |}
			\hline
			Subgroup & Number of PMSG-WTGs \\
			\hline
			&1, 2, 3, 7, 8, 13, 14, 15, 19, 20, 21, 26, 27, 28, 32, 33,\\
Subgroup 1	&34, 35, 38, 39, 40, 44, 45, 46, 50, 51, 52, 56, 57, 58, 62,\\
			&63, 68, 69, 70, 74, 75, 80, 81, 87, 88, 91, 94\\
			\hline
			&4, 5, 6, 9, 10, 11, 12, 16, 17, 18, 22, 23, 24, 25, 29, 30,\\
Subgroup 2	&31, 36, 37, 41, 42, 43, 47, 48, 49, 53, 54, 55, 59, 60, 61,\\
			&64, 65, 66, 67, 71, 72, 73, 76, 77, 78, 79, 82, 83, 84, 85,\\
			&86, 89, 90, 92, 93, 95, 96, 97, 98, 99, 100\\
			\hline
			Subgroup 3&--\\
			\hline			
		\end{tabular}
	}
	\end{center}
\end{table}

\begin{figure}[!htb]
	\centering
	\includegraphics[width=3.5 in]{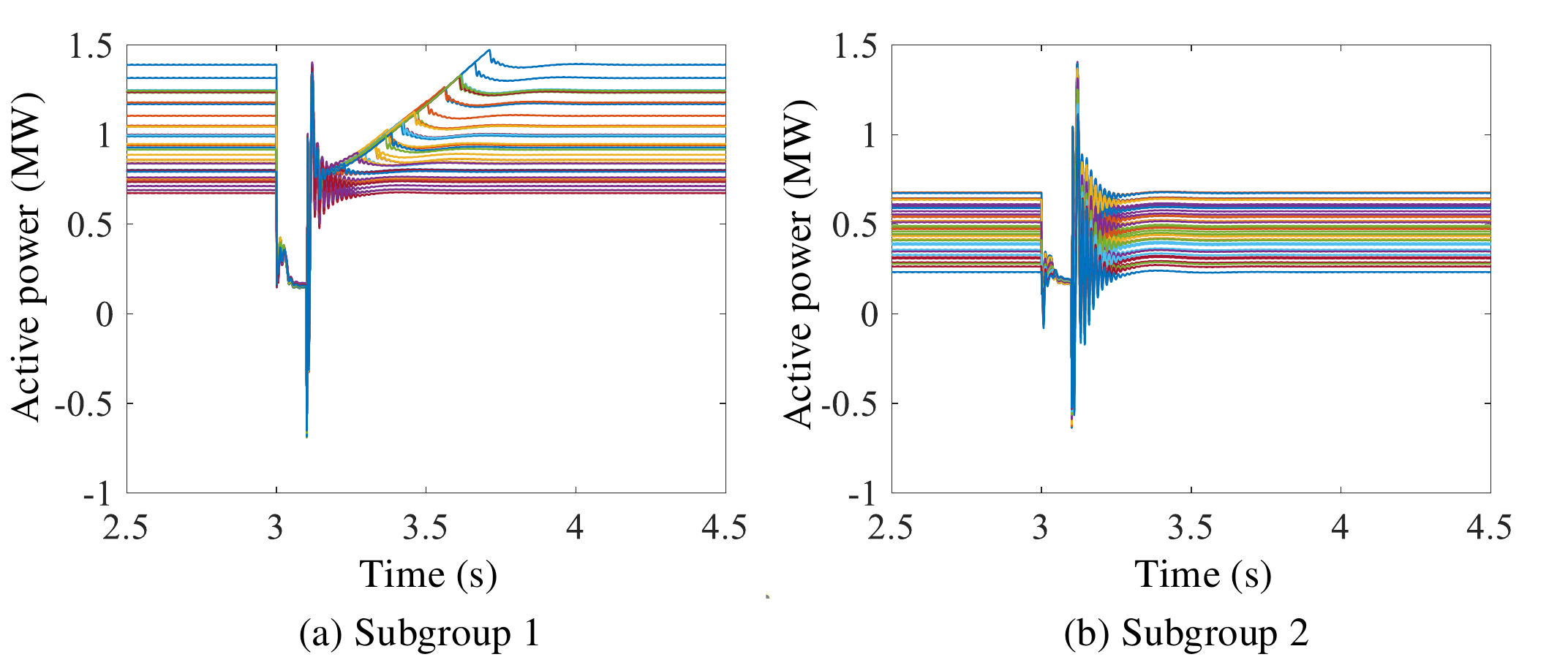}
	\caption{Active power responses of individual PMSG-WTG in the same subgroup in Case I.}
	\label{fig_18}
\end{figure}

\begin{figure}[!htb]
	\centering
\vspace{-0.3cm}  
	\includegraphics[width=2.8 in]{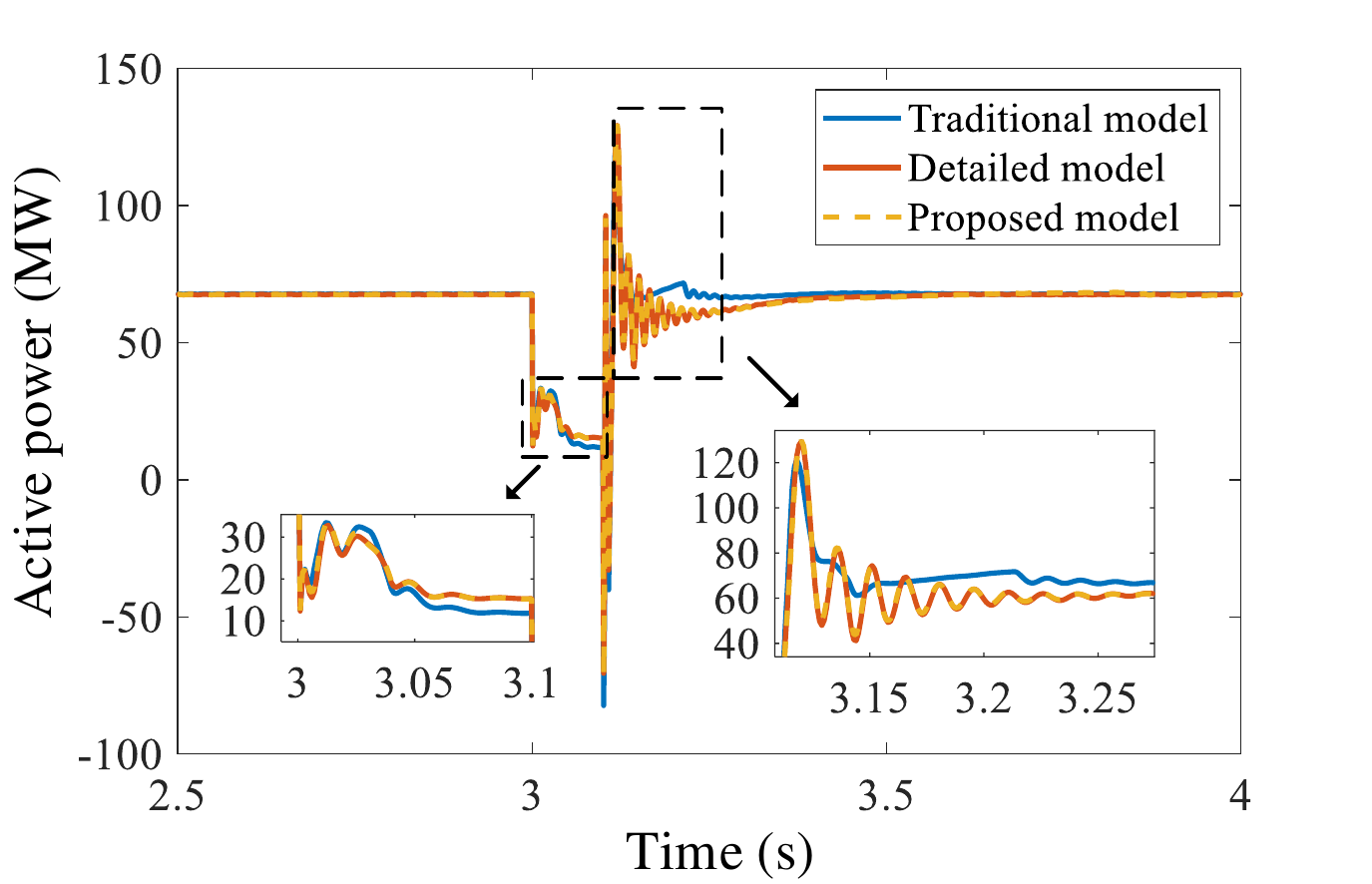}
	\caption{The comparison of active power in Case I.}
	\label{fig_19}
\end{figure}

\begin{figure}[!htb]
	\centering
	\vspace{-0.3cm}  
	\includegraphics[width=2.8 in]{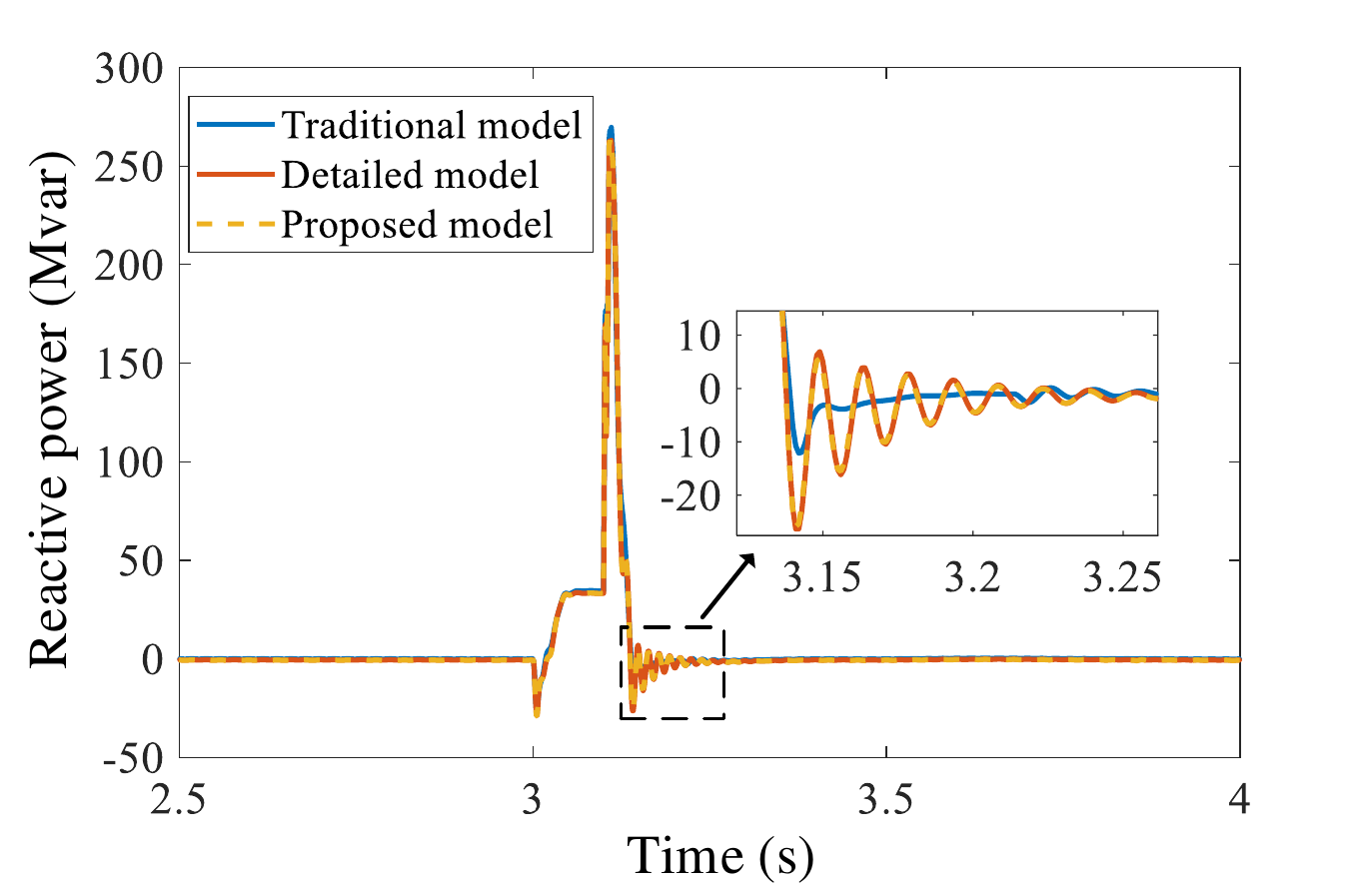}
	\caption{The comparison of reactive power in Case I.}
	\label{fig_20}
\end{figure}

The active power responses of individual PMSG-WTG in the same subgroup are shown in figure. \ref{fig_18}.
It can be found that the response characteristics of the PMSG-WTGs in the same subgroup are basically consistent with each other, which indicates the correctness of the method proposed in Section III. In order to prove the efficiency and correctness of the proposed equivalent method, the dynamic responses of the detailed model, the traditional multi-machine equivalent model based on the wind speed clustering [18] and the proposed equivalent model are presented in Figure. \ref{fig_19} and  Figure. \ref{fig_20}.

The results show that the proposed model is more accurate than the traditional model and can significantly reduce the simulation time compared to the detailed model, as shown in Table \ref{Method_Compare}.
\subsection{Case II:Verification of the Method When a Distant Fault Occurs}
When a three-phase short circuit fault occurs at node 24, the proposed method can be used to calculate the PCC voltage and terminal voltage of each PMSG-WTG, which is similar to the procedure in Case I. The final result of PCC voltage is $\alpha ^{f^{-'}}_{pcc}=0.62$ and the clustering results are shown in Table \ref{Clustering Results of case II.}.

\begin{table}
	\renewcommand\arraystretch{1.2}
	\begin{center}
		\caption{Clustering Results of Case II.}
		\label{Clustering Results of case II.}
		\resizebox{0.99\linewidth}{0.08\textheight}{
		\begin{tabular}{| c | c |}
			\hline
			Subgroup & Number of PMSG-WTGs \\
			\hline
			Subgroup 1&--\\
			\hline 
\multirow{2}{*}{Subgroup 2}&1, 13, 14, 19, 26, 32, 33, 38, 39, 44, 45, 50, 51, 56, 57,\\
			&68, 69, 80\\
			\hline
			&2, 3, 4, 5, 6, 7, 8, 9, 10, 11, 12, 15, 16, 17, 18, 20, 21,\\
			&22, 23, 24, 25, 27, 28, 29, 30, 31, 34, 35, 36, 37, 40, 41,\\
Subgroup 3	&42, 43, 46, 47, 48, 49, 52, 53, 54, 55, 58, 59, 60, 61, 62,\\
			&63, 64, 65, 66, 67, 70, 71, 72, 73, 74, 75, 76, 77, 78, 79,\\
			&81, 82, 83, 84, 85, 86, 87, 88, 89, 90, 91, 92, 93, 94, 95,\\
			&96, 97, 98, 99, 100\\
			\hline			
		\end{tabular}
	}
	\end{center}
\end{table}

\begin{figure}[!htb]
	\centering
	\includegraphics[width=3.5 in]{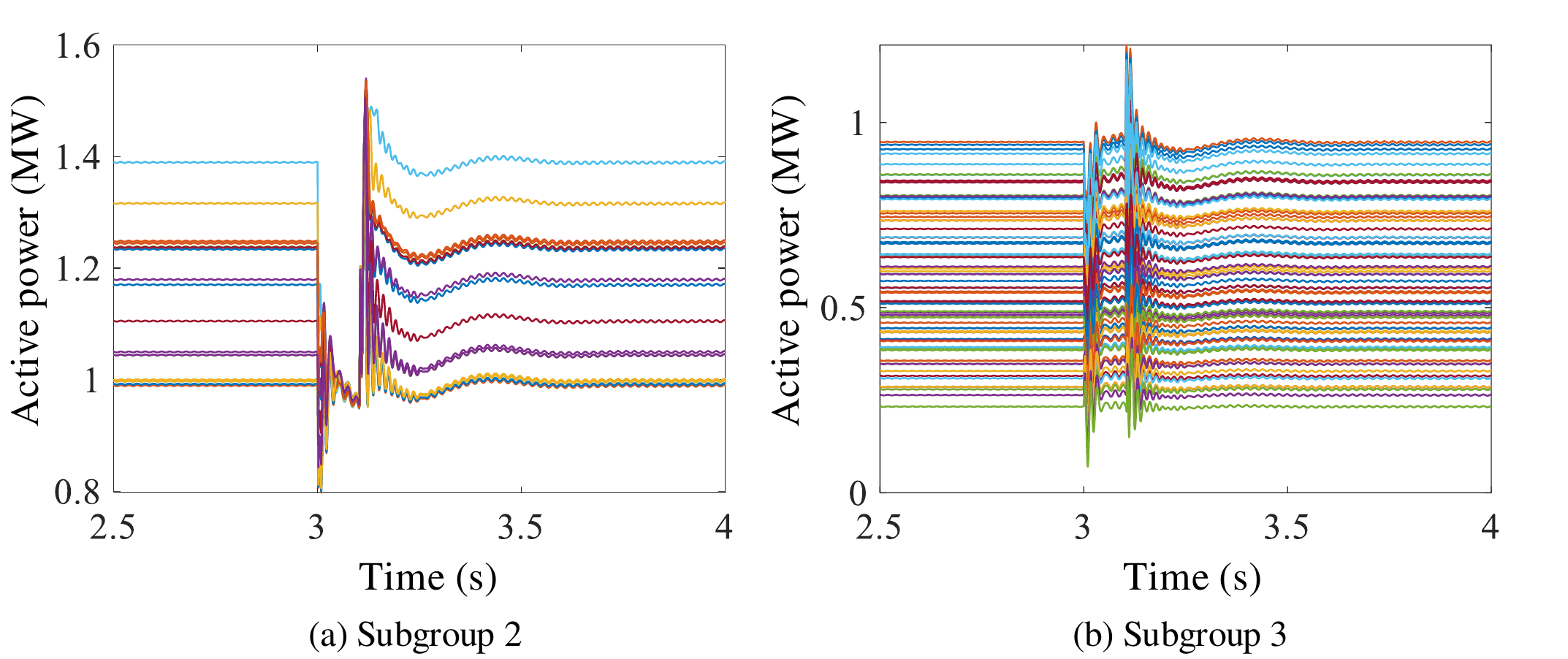}
	\caption{Active power responses of individual PMSG-WTG in the same subgroup in Case II.}
	\label{fig_21}
\end{figure}

The active power responses of individual PMSG-WTG in the same cluster are shown in figure. \ref{fig_21}.

\begin{figure}[!htb]
	\centering
\vspace{-0.3cm}  
	\includegraphics[width=2.8 in]{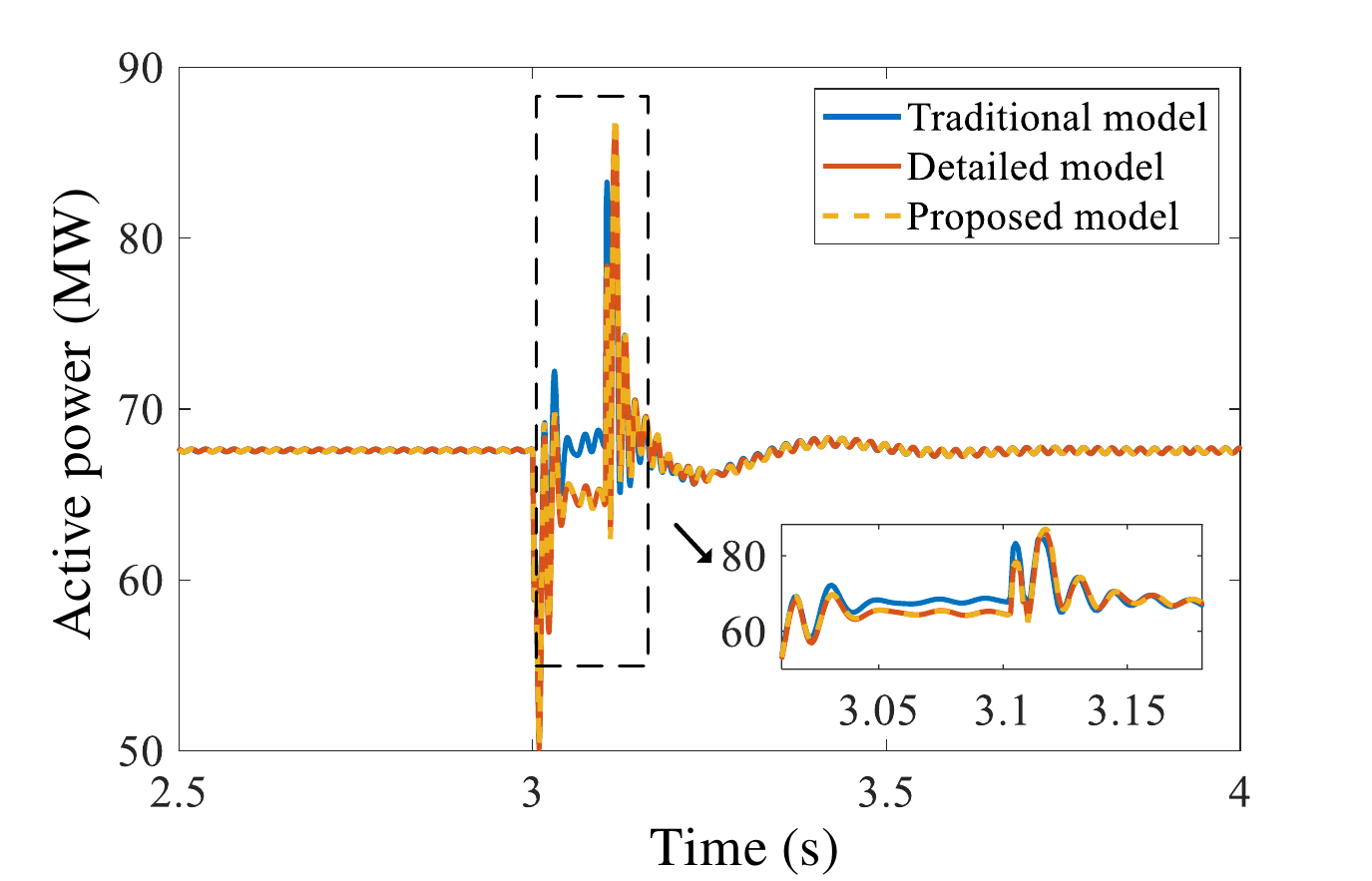}
	\caption{The comparison of active power in Case II.}
	\label{fig_22}
\end{figure}

\begin{figure}[!htb]
	\centering
	\vspace{-0.3cm}  
	\includegraphics[width=2.8 in]{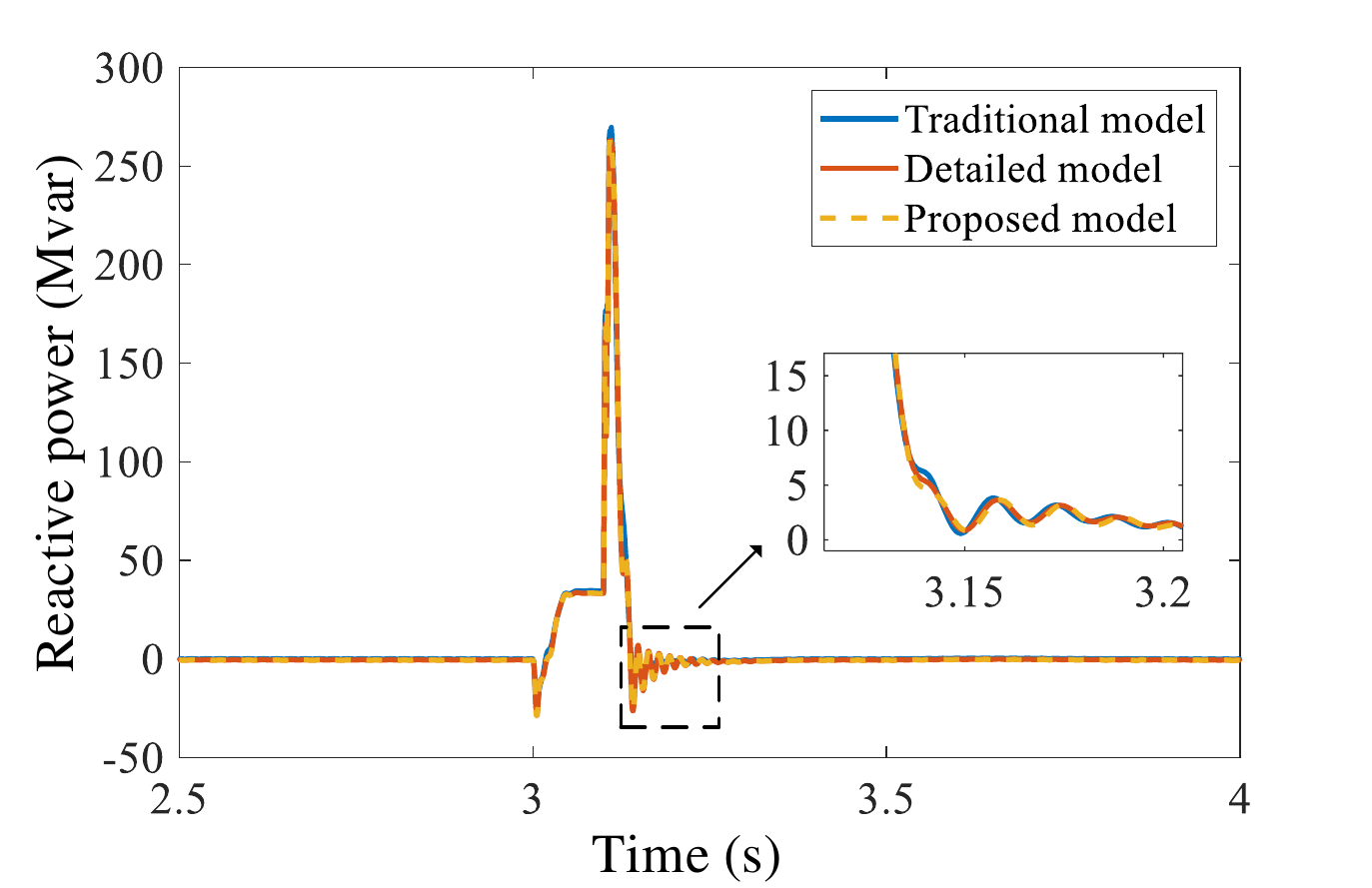}
	\caption{The comparison of reactive power in Case II.}
	\label{fig_23}
\end{figure}

The response characteristics of PMSG-WTGs in subgroup 2 and 3 are consistent with the results of the theoretical analysis in Section III. And the dynamic responses of different models are presented in Figure. \ref{fig_22} and Figure. \ref{fig_23}. The simulation times and  equivalent errors are compared in Table \ref{Method_Compare}. The simulation time of the proposed method is the sum of the time spent in each iteration and the errors are calculated from the mean absolute percentage error between the active responses.

\begin{table}[ht!]
	\centering

	\begin{center}
		\caption{Simulation Times and Equivalent Errors of different methods.}
		\label{Method_Compare}
		\resizebox{0.99\linewidth}{0.07\textheight}{
		\begin{tabular}{| c | c | c | c | c |}
			\hline
			 \multirow{3}{*}{}		& \multicolumn{2}{c}{Case I} \vline& \multicolumn{2}{c}{Case II} \vline\\
			\cline{2-5}
			 \multirow{3}{*}{}& Simulation & Equivalent & Simulation& Equivalent\\
			 \multirow{3}{*}{}& time (s)   &  error (\%) & time (s) & error (\%)\\
			\hline		 		
Traditional	& \multirow{2}{*}{8.29}		&\multirow{2}{*}{7.64}		& \multirow{2}{*}{7.43} &\multirow{2}{*}{0.81}\\
model	& \multirow{2}{*}{}		& \multirow{2}{*}{}					& \multirow{2}{*}{}	 &\multirow{2}{*}{}	\\
			\hline		 		
Proposed	& 4.12+5.22+				&\multirow{2}{*}{0.75}				& 3.72+4.79+ &\multirow{2}{*}{0.14}	\\
model		&10.43=19.77  				&\multirow{2}{*}{}					&10.22=18.73		&\multirow{2}{*}{}	\\
			\hline		 		
detailed	& \multirow{2}{*}{1071.43}		& \multirow{2}{*}{--}		& \multirow{2}{*}{1037.08} &\multirow{2}{*}{--}\\
model	& \multirow{2}{*}{}				& \multirow{2}{*}{}					&\multirow{2}{*}{}&\multirow{2}{*}{}\\
			\hline			
		\end{tabular}}

	\end{center}	
\end{table}

According to case I, when there are some PMSG-WTGs belonging to subgroup I, the traditional equivalent model is less capable of equating the active power response during the fault recovery. In case II, there are some PMSG-WTGs belonging to subgroup III and no PMSG-WTG belonging to subgroup I. The traditional equivalent model can properly reflect the active power response during the fault recovery but is inaccurate during the fault duration. The above two cases illustrate that different fault severity degrees have great impacts on the clustering results. Thus, the equivalent methods without considering the faults are not accurate. As can be seen from Table \ref{Method_Compare}, the proposed method improves the simulation efficiency of the detailed model while substantially increasing the accuracy of the traditional equivalent model, which verifies the effectiveness of the proposed method.

\section{Conclusion}
Based on the analysis of the active transient response characteristics of PMSG-WTGs, a clustering method considering wind speeds and PMSG-WTG terminal voltages is proposed. And a single-machine equivalent method is proposed for each subgroup of PMSG-WTGs. For the subgroup of PMSG-WTGs with active power ramp recovery process, an equivalent model with segmented ramp rate limitation for active current is proposed, which can more accurately characterize the dynamic responses of wind farm during fault recovery. 
The simulation results show that the proposed method greatly improves the simulation efficiency compared with the detailed model. And compared with the multi-machine equivalent method based on the wind speed clustering, the proposed method has greatly improved the accuracy.

\bibliographystyle{IEEEtran}
\bibliography{ref2}
\end{document}